# Degradation of Black Phosphorus: The Role of Oxygen and Water

Yuan Huang,[1,2,3†] Jingsi Qiao,[4†] Kai He,[2] Stoyan Bliznakov,[5] Eli Sutter,[6] Xianjue Chen,[1,3] Da Luo,[1,3]

Fanke Meng,[5] Dong Su,[2] Jeremy Decker,[1] Wei Ji,[4,*] Rodney S. Ruoff,[1, 3*] and Peter Sutter[7,*]

[1]*Center for Multidimensional Carbon Materials (CMCM), Institute for Basic Science (IBS), Ulsan 689-798, Republic of Korea*

[2]*Center for Functional Nanomaterials, Brookhaven National Laboratory, Upton, NY 11973, USA*

[3]*Department of Chemistry and School of Materials Science and Engineering, Ulsan National Institute of Science and Technology, Ulsan 689-798, Republic of Korea*

[4]*Department of Physics and Beijing Key Laboratory of Optoelectronic Functional Materials & Micro-nano Devices, Renmin University of China, Beijing 100872, China*

[5]*Chemistry Department, Brookhaven National Laboratory, Upton, New York 11973, USA*

[6]*Department of Mechanical & Materials Engineering, University of Nebraska-Lincoln, Lincoln, NE 68588, USA*

[7]*Department of Electrical & Computer Engineering, University of Nebraska-Lincoln, Lincoln, NE 68588, USA*

## Abstract

Black phosphorus (BP) has attracted significant interest as a monolayer or few-layer material with extraordinary electrical and optoelectronic properties. However, degradation in air and other environments is an unresolved issue that may limit future applications. In particular the role of different ambient species has remained controversial. Here, we report systematic experiments combined with ab-initio calculations that address the effects of oxygen and water in the degradation of BP. Our results show that BP rapidly degrades whenever oxygen is present, but is unaffected by deaerated (i.e., $O_2$ depleted) water. This behavior is rationalized by oxidation involving a facile dissociative chemisorption of $O_2$, whereas $H_2O$ molecules are weakly physisorbed and do not dissociate on the BP surface. Oxidation (by $O_2$) turns the hydrophobic pristine BP surface progressively hydrophilic. Our results have implications on the development of encapsulation strategies for BP, and open new avenues for exploration of phenomena in aqueous solutions including solution-gating, electrochemistry, and solution-phase approaches for exfoliation, dispersion, and delivery of BP.

†These authors contributed equally to this work. *To whom correspondence should be addressed: Peter Sutter (E-mail: psutter@unl.edu); Rodney S. Ruoff (E-mail: ruofflab@gmail.com); Wei Ji (E-mail: wji@ruc.edu.cn).

Black phosphorus (BP) stands out among the family of two-dimensional materials due to several of its unique properties. BP is a direct bandgap, anisotropic and high mobility (~1000 cm$^2$/V·s) semiconductor with a thickness-dependent fundamental gap varying from 0.3 eV in the bulk to 1.5~2.0 eV for the monolayer,[1-6] which covers large portions of the visible and near infrared electromagnetic spectrum. Black phosphorus also has lower reactivity than most of other elemental 2D semiconductors, for example silicene, and can be handled under ambient conditions,[7] at least for limited time periods.[8] Although its interlayer interaction is stronger than that of graphite, MoS$_2$ and most other layered crystals,[9] mechanical exfoliation allows the preparation of BP samples down to the monolayer limit.[10] The valence band maximum (VBM) of BP lies around -4 to -5 eV,[11, 12] which is energetically higher than the Fermi levels of typical metals used for electric contacts; this leads to BP showing p-type doping independent of thickness.[1, 11] This combination of properties, contrasting with those of semi-metallic graphene,[13] air unstable silicene, and n-type indirect bandgap multilayer MoS$_2$,[7, 14] make BP a highly promising material for future electronic and optoelectronic applications.

A remaining key issue of BP toward these applications, however, lies in its degradation under ambient conditions. Formation of pits and bubbles on bulk BP revealed by scanning tunneling microscopy (STM) was ascribed to electrochemical reactions occurring between the STM tip and the BP surface.[15] Even without an applied bias voltage, droplet-like structures are also observable on the surface of BP exposed to air by atomic force and optical microscopy.[8, 16, 17] These structures are believed to be water droplets since experimental and theoretical studies on the environmental stability of BP suggests that the surface is intrinsically hydrophilic; long-term exposure to air and moisture can completely etch away the material.[8, 18] It was also found that the thinner the BP flakes, the faster the water adsorption,[17] and even boron nitride top-encapsulation appears unable to block the species that degrade BP field effect transistor (FET) devices.[16] However, another explanation suggested that oxygen could be responsible for the degradation of BP and the chemisorbed oxygen atoms could increase the hydrophilicity of the BP surface; this alternative explanation makes the mechanism of BP degradation interesting and potentially complicated.[18] A recent study suggested that water, oxygen and visible light are simultaneously required for the degradation of BP,[19] but the detailed roles of these three parameters were not clearly specified. While all these investigations suggest water as one of the primary species responsible for BP degradation, this conclusion contrasts sharply with the finding that devices, such as humidity sensors comprising BP nano-flakes can be stable during extended time exposures to water without substantial degradation in device performance.[20] In addition, recent work by Kang et al. demonstrated the successful dispersion and delivery of optically and electronically active few-layer BP in deaerated aqueous solutions instead of anhydrous organic solvents.[21] Taken together, these recent reports call into question the assumed role of water in the degradation of black phosphorus.

Here, we carried out a joint experimental and theoretical study with the goal of comprehensively addressing and clarifying the physics and chemistry involved in the breakdown of BP, focusing in particular on the interaction of reactive species in air ($O_2$, $H_2O$, etc.) with BP, their effects on electronic properties and their role in the chemical decomposition of BP. We prepared BP flakes of different thickness – bulk, few-layer, down to monolayer – on SiO$_2$/Si support by a modified mechanical exfoliation method.[22] The evolution of the BP flakes with exposure to air, water with dissolved oxygen, and properly deaerated water was monitored by optical microscopy and tapping-mode atomic force microscopy (AFM), and characterized by (scanning) transmission electron microscopy ((S)TEM) and electron energy loss spectroscopy (EELS). We found that oxygen plays a crucial role in the degradation of BP whereas deaerated water alone does not affect the material. Density functional theory (DFT) calculations reveal the detailed atomic processes of the degradation, showing that H$_2$O physisorbs on pristine BP, but O$_2$ adsorbs dissociatively.[23, 24] The oxidized BP surface doubles the adsorption energy of H$_2$O, making the surface hydrophilic. These findings are further supported by FET and electrochemical measurements, as well as [18]O isotope labeling experiments. Our results, especially the stability of BP in pure water, may promote new directions such as the exploration of the electrochemical properties of BP and liquid dispersion of BP for handling and delivery via aqueous solutions.

**RESULTS AND DISCUSSION**

To test the environmental stability of BP, we prepared flakes of different thickness (bulk, few-layer, and monolayer) on SiO$_2$/Si supports by using a modified exfoliation method,[22] and followed their evolution in air by optical microscopy and AFM (see section S1, Supporting Information). Our findings are consistent with previous reports of the morphology change of BP in air.[8, 25] Focusing on the specific role of water in the chemical reactivity of BP, we immersed exfoliated flakes supported on SiO$_2$ for different time periods in deionized (DI) water, ultimately followed by exposure to air (Figure 1). Figure 1a shows an optical micrograph of a freshly exfoliated BP flake. Subsequently, the sample was immersed in DI water. As shown in Figure 1b, after one week in DI water the morphology of the flake did not show any visible changes, and no droplets were observed in its vicinity. After 2 weeks in DI water, the optical contrast of the flake had evidently changed (Figure 1c), in a way that suggests that the thickness was modified across the entire flake. As we will show below, oxygen dissolved in water is responsible for this etching. The lack of observable features in the vicinity of the flake suggests that the drop-like features frequently observed after air exposure are soluble in or react with water and have been eliminated during immersion, thus supporting the assumption of previous reports that these drops may consist of aqueous decomposition products of BP.[8] On the other hand, this also suggests that the surface of BP may not be intrinsically hydrophilic, as reported by Du *et al.* and Castellanos-Gomez *et al.*,[8, 26] because no droplets were observed even after the BP was taken out of DI water. After the prolonged immersion in DI water, the same sample was exposed to air for 1 week. As seen in Figure 1d, the flake decomposed completely, leaving behind a number of drop-shaped residues of different size within its previous outline. This indicates that the decomposition in air is much faster than in DI water. The sample was subsequently loaded into ultrahigh vacuum (UHV, $10^{-9}$ Torr) for several hours, but the droplet-like morphology did not show any change. Annealing in UHV at 120 ℃ for 2 hours (Figure 1e) led to a slight decrease in the droplet density but the larger droplets still remained. Only after annealing at higher temperature (250 ℃ for 2 hours, Figure 1f) also the large droplets were replaced by drying patterns within their original footprint. These observations suggest that the droplet-like objects do not mainly consist of water, but are accumulations of species with lower vapor pressure, such as phosphorus oxides ($P_xO_y$) or phosphoric acid; thus the degradation mechanism may not be as previously reported.[8, 19] Phosphorus trioxide ($P_2O_3$) and pentoxide ($P_2O_5$) are both solids near room temperature ($P_2O_5$ sublimes; $P_2O_3$ melts at 23.8 °C). $P_2O_3$ has high vapor pressure (7.5 Torr at 47 °C; 75 Torr at 100 °C), whereas $P_2O_5$ is far less volatile (vapor pressure $7.5 \cdot 10^{-3}$ Torr at 285 °C).[27] Hence, our experiments involving vacuum annealing of air-exposed BP are consistent with a reaction of BP with oxygen in air to form a phosphorus oxide with composition close to $P_2O_5$.

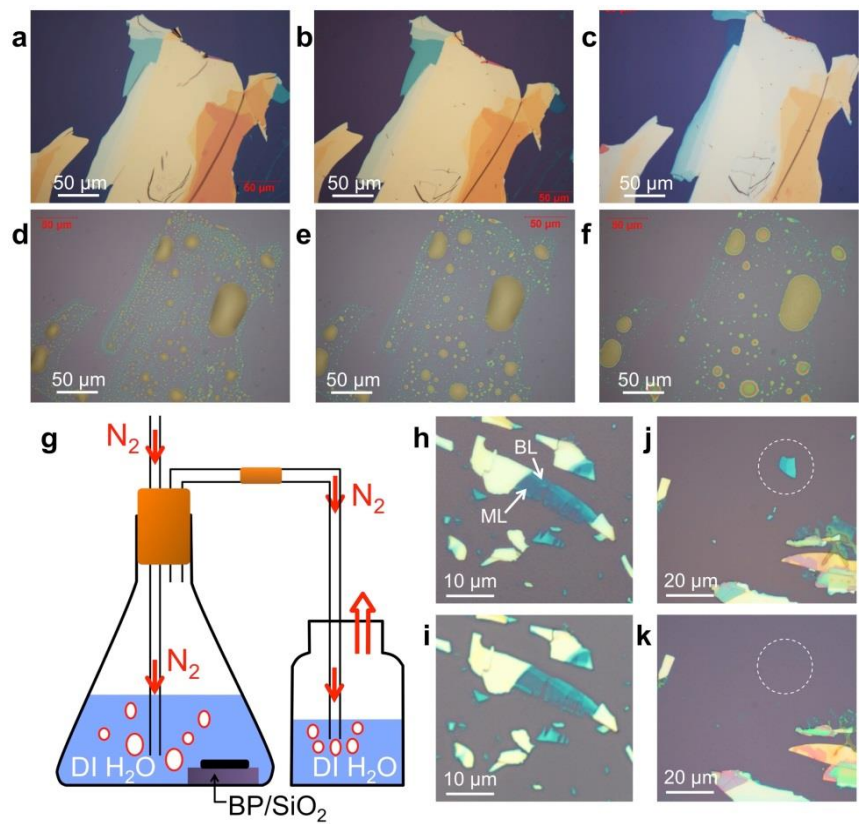

**Figure 1. Morphology change of an exfoliated BP flake on SiO₂/Si after exposure to water and air. a.** Optical images of a freshly exfoliated BP flake. **b.** Same flake after being submerged in DI water for 1 week. **c.** Same flake after a total of 2 weeks exposure to water. **d.** Image after removal of the flake from DI water and exposure to air for 1 week, showing the complete dissolution of the flake and droplet-like residues within its footprint. **e.** Same sample after annealing at 120 °C for 2 hours in ultrahigh vacuum (UHV, $10^{-9}$ Torr). **f.** Sample after additional annealing at 250 °C for 2 hours in UHV. **g.** Schematic diagram of the setup used to remove dissolved $O_2$ from DI water (by bubbling $N_2$) and, alternatively, enriching water with $O_2$ (by bubbling $O_2$). **h.** Optical micrograph of freshly exfoliated BP flakes on SiO₂/Si. Based on their optical contrast, the deep blue segments are identified as monolayer and bilayer BP. **i.** Same sample region after immersion in oxygen-depleted water ($N_2$ bubbling) for 2 days. **j.** Freshly exfoliated BP flakes. **k.** Same sample region after immersion in oxygen-enriched DI water ($O_2$ bubbling) for 2 days.

In order to better understand the decomposition pathway, we compared the degradation process under well-defined conditions: in air and DI water (see section S1, Supporting Information); and in oxygen-depleted (*i.e.*, deaerated) and oxygen-enriched DI water (Figure 1h, i and Figure 1j, k, respectively). The comparison with the sample exposed to air shows that decomposition of BP is substantially slower in DI water, but some residual reactivity remained. It is well known that molecular oxygen can be dissolved in water. At 25 °C, the equilibrium solubility of $O_2$ in water is 8.27 mg/L (1.26 mM), but by deaeration the concentration of $O_2$ can be reduced by 3-4 orders of magnitude (to $\sim 10^{-7}$ M). To control the oxygen concentration in water, we used a setup in which different gases could be controllably bubbled through the DI water reservoir in which the BP sample was held (Figure 1g). To lower the concentration of dissolved $O_2$, nitrogen gas was bubbled continuously through the DI water. A sample containing flakes with both monolayer and bilayer BP was exposed to these conditions for 2 days, after which the flake remained almost completely unchanged (Figure 1i). Notably, the monolayer and bilayer areas are still observable, *i.e.*, have not been affected as in the case of neat DI water with a higher (equilibrium) concentration of dissolved $O_2$. Some groups claimed that thinner flakes show accelerated decomposition,[17, 19] yet under oxygen-depleted conditions the BP flakes remained stable down to monolayer thickness. This suggests that BP is unable to activate and react with $H_2O$ itself and that the primary reaction pathway requires $O_2$, either present in air or dissolved in water. To further test the apparent key role of $O_2$, we performed a control experiment in which the water was enriched in dissolved oxygen by bubbling $O_2$ instead of $N_2$. As can be seen by comparing Figures 1j and k, even few-layer flakes are completely etched away under these conditions.

Figure 2a shows a high-angle annular dark-field STEM (HAADF-STEM) image of the pristine (freshly exfoliated) BP flakes on amorphous carbon support. High-resolution TEM (HRTEM) imaging reveals the atomic structure of a BP flake in plan-view (along the [110] axis, Figure 2b) and cross-section (Figure 2c), showing that the flake is a single crystal with interlayer spacing of ~5.4 Å. In order to monitor structural and chemical changes due to ambient exposure, we performed electron energy loss spectroscopy in STEM (STEM-EELS) within the same region (red rectangle in Figure 2a) before and after exposing the sample to air for 1 day. Comparing the HAADF-STEM image contrast in Figure 2d (pristine) and Figure 2e (after 1 day in air), we found the thickness of the flake was reduced after air exposure.

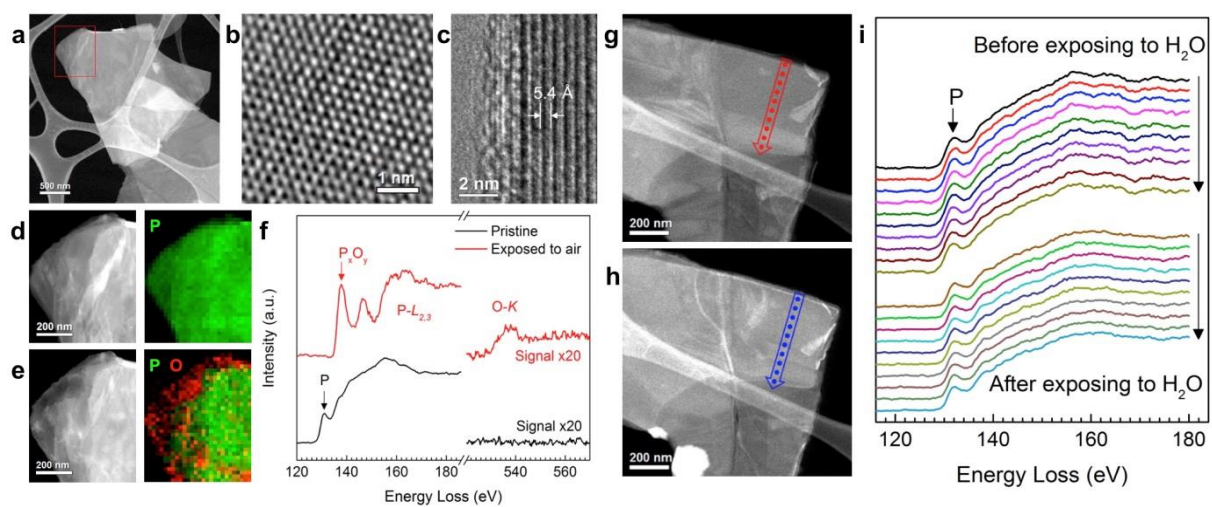

**Figure 2. (S)TEM and EELS characterization of a multilayer BP flake before and after air and water exposure. a.** HAADF-STEM image at 200 kV of a flake immediately after exfoliation and transferring onto a TEM grid. **b.** High-resolution TEM image of a BP flake. **c.** Cross-sectional TEM image of a folded section of the flake. The basal plane is along the [110] axis. The interlayer spacing is ~5.4 Å. **d.** Enlarged STEM image within the rectangular area marked in a, and corresponding 2D STEM-EELS map of the BP flake immediately after exfoliation. **e.** STEM image and 2D STEM-EELS map after exposure to air for 1 day. P is shown in green, and O in red. **f.** Representative EELS spectra of pristine and oxidized (1 day air exposure) BP samples. **g.** HAADF-STEM image of a freshly exfoliated BP flake. **h.** Same flake after submersion in DI water for 1 day. **i.** EELS linescan profiles along the arrows in g and h, showing the phosphorus L-edge without any detectable oxidization.

The EELS spectrum taken from the darker (thinner) region is compared to that of the pristine sample in Figure 2f. The characteristic energy-loss edges, the $L_{2,3}$-edge of P and the K-edge of O, can be used to identify the chemical valance states. Our data clearly show that the phosphorus $L_{2,3}$-edge shifts from 130 eV to 136 eV and the oxygen K-edge at 532 eV appears after exposure to air, both indicating the oxidation of P to form $P_xO_y$. The elemental distribution of P and O in the observed region can be revealed by false-color STEM-EELS maps (Figure 2d, e). Combining the STEM and EELS results, we find a clear correlation between the morphology change (thickness reduction) and the local chemical change (oxidation).

The same type of characterization was performed on a similar BP flake before and after immersion in water for 1 day. HAADF-STEM images obtained before (Figure 2g) and after (Figure 2h) exposure to $H_2O$ do not show any obvious changes in morphology. The corresponding energy-loss profiles are displayed in Figure 2i. The spectra do not show any significant chemical shift, but show phosphorus remaining mostly in the elemental (zero-valent $P^0$) state. There is a small, but detectable increase in intensity at ~138 eV, indicating the formation of traces of $P_xO_y$ likely by reaction with oxygen during the sample transfer and loading process. Comparing the results after air and water exposure, we find ratios of P $L_{2,3}$-edge intensities of $P^0$:$P_xO_y$ of 0.24 (1 day in air) and 2.3 (1 day in DI water). This strongly suggests that crystalline BP is unable to oxidize via dissociation of water, $e.g.$, via reactions such as but instead is oxidized exclusively by reaction with $O_2$, either present in air or dissolved in water. This means that even though bulk red phosphorus and white phosphorus compounds can react with water,[28-31] crystalline black phosphorus is evidently more stable against reaction with water than these other P allotropes. Besides water, BP also remains stable in other aqueous solutions, including high concentration HCl and KOH, as discussed in Supporting Information S4.

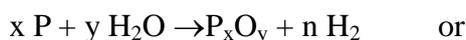

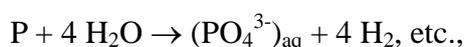

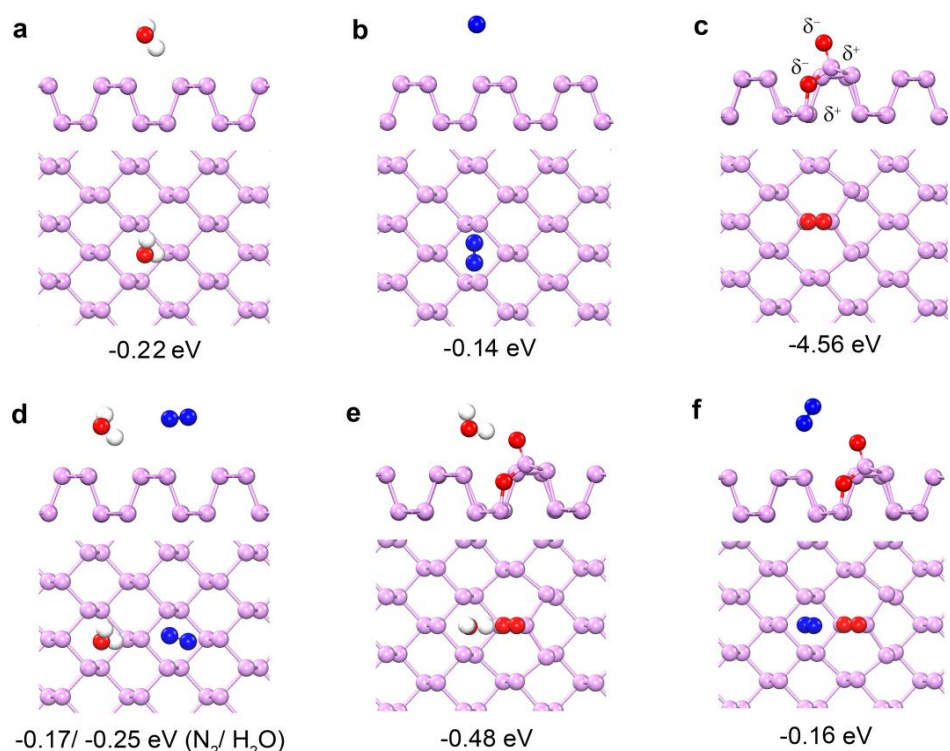

**Figure 3. Physisorption and chemisorption of $H_2O$, $N_2$ and $O_2$ on BP.** Atomic structures of physisorbed $H_2O$ (**a**) and $N_2$ (**b**) on pristine monolayer (1L) BP. At room temperature, $O_2$ molecules dissociatively attach to 1L BP and the most stable configuration of IL-2O shown in **c**, which one Oxygen bonds to the lone pair and the other one sits in-between two P atoms. **d**. The joint adsorption of $H_2O$ and $N_2$ on pristine 1L BP. **e**, **f**. $H_2O$ (**e**) and $N_2$ (**f**) individually adsorbed on oxidized 1L BP surface, respectively.

DFT calculations were carried out to understand the degradation process of BP from an atomistic point of view. We have considered interaction with oxygen, water and nitrogen molecules, the primary constituents of air. All these molecules could physisorb on the BP surface, with the most favorable adsorption sites of $H_2O$ and $N_2$ shown in Figure 3a and b, respectively. It appears that $H_2O$ interacts more strongly with the surface than either $O_2$ or $N_2$ as evidenced by the 0.03 or 0.08 eV stronger interaction energy. Although the interaction energy of $H_2O$-BP is stronger, it is still 4 meV smaller than the intermolecular interaction energy of water (-0.222 eV), indicating that BP is hydrophobic. How can these weakly interacting molecules drastically affect the physical properties of BP? We found that the physisorption state of $O_2$ is actually only a metastable state, and $O_2$ shows dissociative adsorption (DA) on the BP surface forming two dissociated O atoms.[23, 24, 32] The DA attachment is highly exothermic with energy release of over 4 eV per $O_2$ molecule and an activation barrier of 0.62 eV. Such a low barrier allows the DA to occur at room temperature. The most energetically preferred dissociated configuration (adsorption energy -4.56 eV) is shown in Figure 3c. We denote this configuration, in which one oxygen atom binds at the "Interstitial bridge" (IB) site and the other at the "Lone-pair" (LP) site, as configuration **IL**. Calculations of the adsorption energy for a single oxygen atom suggest that the **LP** configuration, in which the O-orbital directly interacts with the lone electron pair of a surface P atom (Figure S11a), is the most stable one; in the next favored configuration, the O atom occupies the **IB** site in-between two P atoms belonging to the upper and lower sub-layers (Figure S11b), respectively. The $O_2$ DA configuration **IL** (Figure 3c) has a 0.15 eV lower energy than the **LL** (**LP**+**LP**) configuration reported in Ref. 18.  The slightly lower energy is ascribed to electrostatic attraction in **IL** and repulsion in **LL**, which override the energy difference between chemisorbed single **LP** and **IB** oxygen. In light of the DA attachment of $O_2$, we examined the possibility that water molecules might adsorb dissociatively on BP. Our results show that at least 1.05 eV is required to break a H-O bond in a physisorbed $H_2O$ molecule, which makes this process energetically unfavorable. We therefore conclude that $O_2$ chemisorbs on the pristine BP surface at room temperature but water only physisorbs as an intact $H_2O$ molecule.

Liquid droplets are often seen on degraded BP surfaces, and usually ascribed to water droplets. The degradation was thus explained based on the assumed strong interaction of water with the BP surface. However, we have shown here that water only weakly physisorbs on the BP surface, implying that the likely direct role of water in degrading the electrical and other properties is minor. Figure 3d shows the joint adsorption of $H_2O$ and $N_2$ on monolayer BP. Figures 3e and 3f show the configurations of $H_2O$ and $N_2$ individually adsorbed on the oxidized BP surface, respectively. We found $N_2$ is so inert that its adsorption energy is only slightly affected (by 0.04 or 0.02 eV) by co-adsorption with $H_2O$ or adsorption on pre-oxidized BP. In contrast, the interaction energy of $H_2O$ is strongly affected by BP surface oxidation. Figure 3e illustrates the most stable configuration of water adsorbed on oxidized BP, in which $H_2O$ forms one hydrogen bond with the **LP** oxygen. Formation of a H-bond in this configuration increases the $H_2O$ adsorption energy to -0.48 eV, roughly 2.2 times larger than on the pristine BP surface (-0.22 eV). This water-surface interaction energy is comparable with a value found previously (-0.49 eV), which was interpreted to represent a superhydrophilic surface.[33] Given a linear relationship between contact angle ($\Theta$) and interaction energy ($E_{ad}$), namely $\Theta = (1-|E_{ad}|/|E_{shp}|) \times 180°$, where $|E_{shp}| = 0.49$ eV is the presumed interaction energy of the superhydrophilic surface (i.e., $\Theta = 0°$), the contact angle is reduced due to oxidation of the surface, from 99° of pristine BP to roughly 4° (nearly superhydrophilic) for IL oxidized BP. All these results manifest that BP oxidation gradually turns the intrinsically hydrophobic surface into a nearly superhydrophilic form. Within this picture, the droplets on BP surface after exposure to air observed in many experiments represent water concentrated in the most strongly oxidized (i.e., superhydrophilic) areas, surrounded by more hydrophobic (i.e., less oxidized) regions of the BP surface.

The adsorption of molecules on the pristine BP surface may affect the intrinsic electronic structure of BP and thus modify the electrical and optical properties of the material. Physisorbed molecules are believed to have minor influence on the electronic structure, which is true for physisorbed $H_2O$ and $N_2$ since all valence and conduction bands of $H_2O$- or $N_2$-covered BP close to the bandgap are nearly unchanged upon adsorption of $H_2O$, as shown in Figure 4a. However, this is not the case for dissociatively adsorbed $O_2$. The two chemisorbed O atoms of the oxidized surface induce a strong charge transfer that substantially modifies the electronic structure (Figure 4b). The oxidized crystal (IL-2O-BP, Figure 3c) has a slightly larger bandgap and increased effective mass. The enlarged effective masses implies a reduced carrier mobility in terms of a phonon-limited picture, in other words oxidation at least partially accounts for the observed degradation of transport properties. Beyond phonon scattering, impurity scatting may also contribute to the reduced carrier mobility. Figure 4b illustrates that the additional adsorption of $H_2O$ or $N_2$ molecules does not appreciably change the electronic structure of oxidized BP, indicating the primary role of the O impurities. Figures 4c and 4d visualize the electronic states of water-covered pristine and oxidized BP surfaces. For $H_2O$ on pristine BP, the valence and conduction bands are continuous across the adsorption site, see Figure 4c. However, oxidization of the BP surface causes the wave functions of valence and conduction bands to become spatially discontinuous around the P atoms in the vicinity of the O-P bonds for **IL**-oxidized BP (Figure 4d). Our combined results compellingly indicate that $O_2$, not $H_2O$ or $N_2$ molecules, through their dissociative adsorption on the BP surface play a dominant role in the degradation of the electronic properties of BP after exposure to air, and modify the surface from hydrophobic to nearly superhydrophilic.

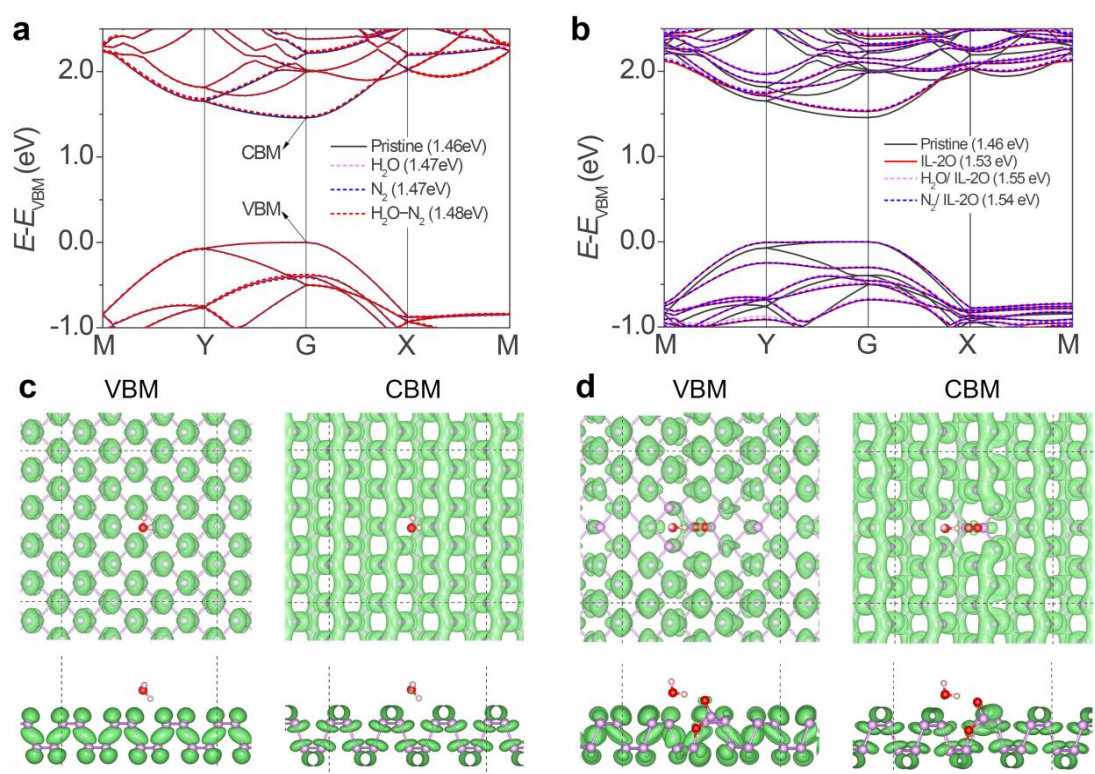

**Figure 4. Effects of adsorption on the electronic structure of BP. a.** Band structures of pristine BP (black solid line), and $H_2O$ (pink dashed line), $N_2$ (blue dashed line) and $H_2O$-$N_2$ (red dashed line) adsorbed on pristine 1L BP. Valence band maximum (VBM) and conduction band minimum (CBM) are marked in the panel. **b.** Band structures of pristine (black solid line) and oxidized BP surfaces (IL-2O, red solid line), and $H_2O$ ($H_2O$/IL-2O, pink dashed line) and $N_2$ ($N_2$/IL-2O, blue dashed line) adsorbed on oxidized BP surfaces. **c, d.** Spatial structures of wave functions for VBM and CBM of $H_2O$ on pristine 1L BP (**c**) and oxidized 1L BP (**d**).

To further test the theoretical results on the wetting of BP by water, isotope labeling experiments were carried out by systemically using $^{18}O_2$ and $H_2^{18}O$ (see Section S8, Supporting Information). Time-of-flight secondary ion mass spectrometry (TOF-SIMS) was performed after one day of exposure of BP flakes to a $^{18}O_2/H_2^{16}O$ mixture prepared by enriching previously deaerated DI water with $^{18}O_2$. TOF-SIMS maps (Fig. S15) show increased concentration of $^{16}$O (originating from adsorbed $H_2^{16}$O) in the same regions that also show high $^{18}$O intensity (due to oxidation by reaction with $^{18}$O$_2$). This finding demonstrates that oxidation makes the BP surface more hydrophilic. In addition, we examined the exposure of freshly cleaved BP to deaerated, isotope labeled water ($H_2^{18}$O), taking precautions to limit its oxidation by contact with air. Here, TOF-SIMS showed no increased $^{18}$O signal, which implies that no detectable accumulation of $H_2^{18}$O took place, i.e., the fresh BP surface is hydrophobic.

In an attempt to determine the effects of air and water on BP devices, we fabricated FETs from exfoliated BP flakes. BP FET devices measured in air showed ambipolar behavior (see Section S3, Supporting Information). The extracted field effect mobility is ~230 cm$^2$/V·s for holes and ~30 cm$^2$/V·s for electrons, respectively, consistent with values observed previously.[3] When measuring the conductance as a function to the time of exposure to air, a brief initial conductance rise was observed that may be due to current annealing;[34] subsequently, the conductance decreased continuously during air exposure, and the device ultimately broke down after an exposure time of ~70 h (Figure S5).

Previous experiments on MoS$_2$ and SnS$_2$ transistors have shown that top gating by high-k dielectrics, such as HfO$_2$ or DI water, among others, can effectively screen scattering centers and thus lead to increased carrier mobility while maintaining high on-off ratios.[35, 36] The use of DI water as an electrolytic top gate can serve two purposes: i) identification if screening by a high-k dielectric (here H$_2$O, with ε ~80 ε$_0$) can lead to increased carrier mobilities in BP, similar to layered metal dichalcogenides; and ii) exploration of the electrical behavior of BP FETs in contact with water. A schematic of the device layout and an optical micrograph of an actual device are shown in Figure 5a. Figure 5b compares $I_{SD}$-$V_{SD}$ characteristics before and after introduction of the water drop used for top gating; the data shown in Figure 5b were measured for zero back-gate and top-gate voltages. Bringing the device channel in contact with DI water substantially increased the conductance. The drain current, $I_{SD}$, at $V_{SD} = 0.2$ V, for example, increased from 9 μA (dry) to 41 μA after applying a drop of DI water (Figure 5b). This implies that even for zero top-gate voltage the DI water either induced a doping of additional charge carriers in the BP channel or reduced impurity scattering due to dielectric screening. Indeed, analysis of the transfer curves with DI water top gate shows dramatic mobility enhancements to ~1260 cm$^2$/V·s in solution-gated BP devices using a DI water drop as gate dielectric. This finding indicates that the conductance increase upon surrounding the device channel with water is primarily due to an increased carrier mobility as a result of effective dielectric screening.

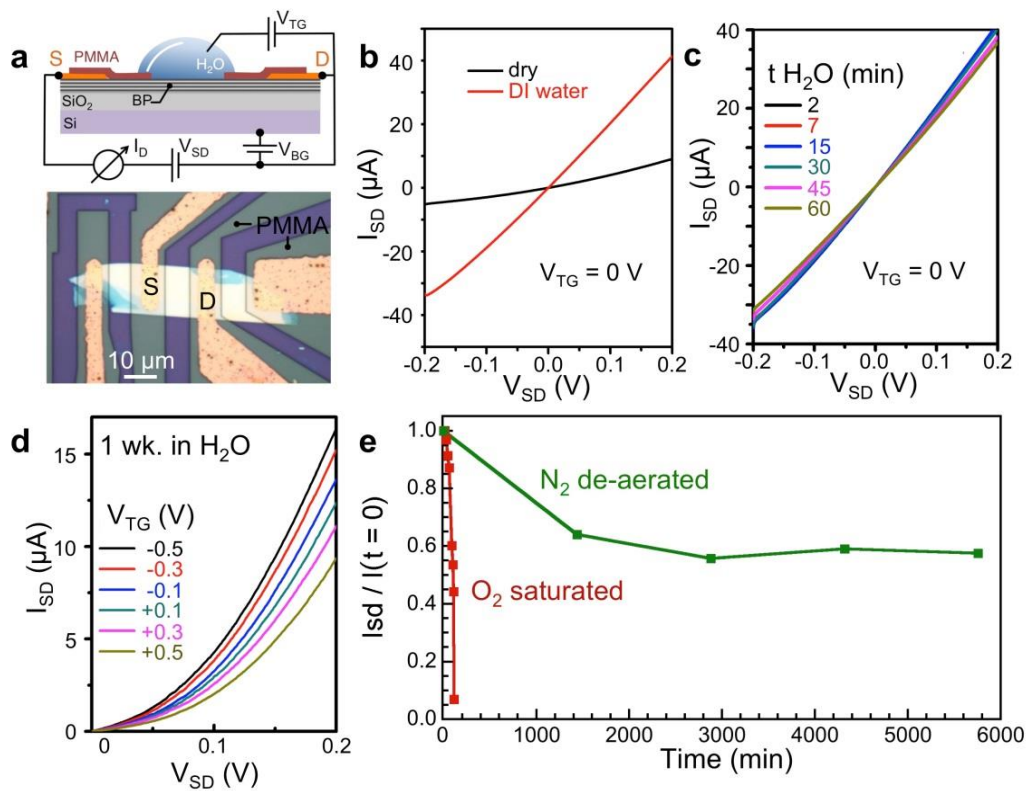

**Figure 5. Electrical transport in a BP FET device with DI water top gate. a.** Schematic of the device geometry, and optical micrograph of the actual FET device with PMMA-covered contacts for top gating by a DI water drop. **b.** $I_{SD}$-$V_{SD}$ characteristics of the BP device without and with DI water top gate (for $V_{TG}$ = 0 V). **c.** $I_{SD}$-$V_{SD}$ characteristics as a function of exposure time of the BP channel to DI water, showing minimal changes over a period of 1 hour. **d.** $I_{SD}$-$V_{SD}$ characteristics for the device after 1 week in contact with water, at different top-gate bias varying from -0.5 V to +0.5 V. **e.** Stability of BP FET devices in $O_2$ saturated and $N_2$ deaerated water.

The stability of the device in contact with water was monitored initially over a period of 1 hour, as shown in Figure 5c. The $I_{SD}$-$V_{SD}$ curves were linear and symmetric for all times, and even though the conductance decreased minimally over 1 hour the characteristics of the device remained nearly unchanged. This result based on FET device characteristics confirms our previous conclusions from exposure to different ambients, namely that BP shows minimal degradation when in contact with DI water. Device measurements also confirm the crucial importance of oxygen in the degradation chemistry. Figure 5e shows a comparison of the evolution of $I_{SD}$ with time for devices kept in water with equilibrium oxygen concentration at room temperature, compared with storage in water that was maintained in a deaerated state by bubbling $N_2$ gas. Whereas $I_{SD}$ for the device held in water with dissolved $O_2$ dropped to near-zero over the course of ~2 hours, the device held in deaerated water showed an initial drop in $I_{SD}$ to about 60% of its original value within ~20 hours, but then maintained a stable conductance with negligible further decrease in $I_{SD}$ for the entire duration of the experiment (~100 hours). Strikingly, the degradation in $O_2$ containing water was even faster than in air. From the absence of significant changes in deaerated water, we conclude that the degradation of BP is dependent on its exposure to oxygen (*i.e.*, the reaction of P $\rightarrow$ P$_x$O$_y$). However, in our experiments with oxygen-containing solutions, water appears to play an important secondary role: it enables the subsequent reaction of P$_x$O$_y$ to the final products, *e.g.*, (PO$_4^{3-}$) ions that dissolve into the solution and form phosphoric acid ($H_3PO_4$). This process ensures the removal of $P_xO_y$ and continued exposure of fresh $P^0$ at the surface and in this way contributes to fast overall reaction rates, hence the faster degradation in oxygen-containing water than in air.

To further evaluate the proposed reaction sequence of the degradation of BP, we performed electrochemical experiments in which redox reactions were driven by electrochemical bias (see Section S5, Supporting Information). Our result suggests that a $P_xO_y$ film is formed on the surface of the BP for potentials ~ +0.4 V. At more positive potentials (higher than +0.5 V), the phosphorus is further oxidizing to $P^{5+}$, forming $(PO_4^{3-})$ ions that dissolve into the solution and form the end product, phosphoric acid ($H_3PO_4$). These observations are consistent with our EELS results, as well as recently reported results by Wang *et al.*[37] The oxidation peak at +0.4 V is not observed in CV curves measured in deaerated solution, in which the reverse potential sweep is stopped before reaching the threshold for the hydrogen evolution reaction (HER, -0.9 V). This indicates that hydrogen gas generated at potentials more negative than -1.0 V chemically reduces the $P_xO_y$ on the electrode surface to elemental phosphorus.[38]

**CONCLUSIONS**

In summary, the degradation of BP under different conditions was systematically characterized structurally, spectroscopically, in device measurements, and via DFT calculations. By comparing exposures to air, water with dissolved oxygen, and deaerated water, we find that the degradation of BP in ambient requires contact with oxygen, and that water does not play a primary role in the reaction, in contrast to previous conclusions. Exposure to water alone does not decompose BP, as is shown by much slower degradation rates in deaerated water compared to the process in air.

DFT calculations reveal that the dissociative adsorption of $O_2$ induces changes in the electronic structure of BP and thus plays an important role in the degradation of the electronic properties. $H_2O$ alone (without $O_2$) cannot cause the degradation of BP since it only physisorbs on the pristine BP surface. Finally, the calculations provide insight into the wetting properties of BP, suggesting that pristine BP interacts weakly with water (i.e., is hydrophobic) whereas oxidation by reaction with $O_2$ gradually turns the surface superhydrophilic. Isotope labeling experiments combined with TOF-SIMS mapping directly demonstrate these predicted changes in the wetting properties of BP due to oxidation.

Our combined results, which demonstrate that BP can be preserved in environments without $O_2$ to suppress degradation, have implications for the identification of suitable encapsulation strategies for BP (e.g., in device applications) by suggesting that the most important role of an encapsulation material is to prevent the contact of BP with oxygen, either directly or via diffusion.

**Methods**

*Sample Preparation*:   Bulk black phosphorous was purchased from HQ-graphene. Adhesive tape was used to peel off a small piece of BP from bulk crystal, and folded 2-3 times to make the flake thinner. The tape with BP was put onto the surface of 300 nm $SiO_2$/Si substrate. All the substrates were cleaned by oxygen plasma to remove organic residual before contact with BP, which improved the yield. Before peeling off, the substrate together with BP/tape was put on a laboratory hot plate to anneal for 1 min at 80 ℃ in order to improve the contact area between BP and $SiO_2$ surface.[22] For the TEM samples, we first spin-coated a layer of photoresist (S1811) on $SiO_2$/Si for 1 min (3000 revolutions per minute (rpm)), and baked it at 100 ℃ for 2 min. Following by exfoliation of BP onto this photoresist surface and checking the flakes under optical microscope, TEM grids were put in contact top with the area containing BP flakes. By dissolving the photoresist in acetone, BP flakes were transferred onto the TEM grid.

*AFM measurements*: Surface morphology and thickness measurements were performed using a Veeco (multimode V) atomic force microscope. All AFM images were collected in tapping mode in order to minimize the degradation by contact with the flakes. AFM imaging was performed in ambient at room temperature.

*TEM-EELS measurement*: The exfoliated flakes were characterized in a JEOL 2100F analytical (scanning) transmission electron microscope ((S)TEM) operated at 200 kV. Samples were kept in acetone after transfer onto TEM grids from photoresist, and quickly loaded in the high-vacuum of the TEM after removing them from acetone.

*Device Fabrication and Electrical Measurements*: After transfer of BP flakes from tape to $SiO_2$/Si substrates by mechanical exfoliation (as described above), the positions of selected flakes were recorded in optical microscopy. A layer of photoresist was spin-coated onto the substrates (S1811, 3000 rpm, 1 min), then put on a hot plate and annealed at 110 ℃ for 2 min. A mask aligner (Karl Suss, MA6) was used to expose the patterns, followed by development in solvent. Ti/Au metal layers (10 nm/50 nm) were deposited by e-beam evaporation. Annealing was carried out in ultrahigh vacuum ($10^{-9}$ Torr) in order to make good contact at the metal/BP interface. Electrical measurements were performed on a four-probe station (Signatone). For DI-water top-gating experiments, we first fabricated conventional (back-gated) FET devices on 300 nm $SiO_2$/Si substrates, and then patterned Poly(methyl methacrylate) (PMMA) windows using electron beam lithography (EBL). The latter provided a defined contact area between the DI water gate and the device channel, while preventing its contact with the metallic source- and drain electrodes.

*Density functional theory (DFT) calculations.* DFT calculations were performed using the generalized gradient approximation for the exchange-correlation potential, the projector augmented wave method,[39, 40] and a plane wave basis set as implemented in the Vienna *ab-initio* simulation package (VASP).[41] The energy cutoff for the plane-wave basis was set to 400 eV for all calculations. A $k$-mesh of 4×4×1 was adopted to sample the first Brillouin zone of the 3×4 monolayer black phosphorus supercell (armchair direction ($x$) and zigzag direction ($y$)). In geometry optimization, van der Waals interactions were considered at the van der Waals density functional (vdW-DF)[42, 43] level with the optB86b functional for the exchange potential (optB86b-vdW)[44, 45]. The shape and volume of each supercell were fixed and all atoms in it were allowed to relax until the residual force per atom was less than 0.01 eV·Å$^{-1}$. Electronic band structures were calculated using the hybrid Heyd-Scuseria-Ernzerhof functional (HSE06) based on the atomic structures optimized by optB86b-vdW. [46, 47] The potential-energy profiles along the reaction paths from physisorbed $O_2$ to dissociative-chemisorbed $O_2$ on 1L prinstine BP were recalculated using the climbing image nudged elastic band (CI-NEB) technique,[48] which locates the saddle points of the reactions, using 8 images.

*Theoretically considered configurations*: Five categories of BP adsorption surfaces were considered in this work, namely physisorbed $O_2$, $N_2$ and $H_2O$ and dissociatively chemisorbed $O_2$ and $H_2O$. Different configurations of each category were proposed according to molecular orientations, directions and sites. Thus, we have a total of 25 initial configurations for $H_2O$, 11 for $N_2$ and 18 for $O_2$; which of $H_2O$ are shown in Figure S13. The DA configurations are similar to the physisorbed configurations and the O, H, or OH atoms/groups bind to the nearest P atoms from their physisorbed configurations. The case of single oxygen adsorption, for example, has five specified sites, namely Lone Pair (LP), Interstitial Bridge O (IB), Dimer Bridge O (DiB-O), Horizontal Bridge (HB) and Diagonal Bridge O (DgB) shown in Figure S11 and the corresponding bandstructures are shown in Figure S12.

**Acknowledgements**

This research used resources of the Center for Functional Nanomaterials, which is a U.S. DOE Office of Science Facility, at Brookhaven National Laboratory under Contract No. DE-SC0012704. Work done at the Center for Multidimensional Carbon Materials was supported by IBS-R019-D1. Work done in Beijing was financially supported by the Ministry of Science and Technology (MOST) of China under Grant No. 2012CB932704, the National Natural Science Foundation of China (NSFC) under Grant Nos. 11274380, 91433103, and the Fundamental Research Funds for the Central Universities, and the Research Funds of Renmin University of China under Grant No. 16XNH062. The authors would like to thank Mingzhao Liu for use of his electrochemical characterization facility. Work at the University of Nebraska-Lincoln was supported by the U.S. Department of Energy, Office of Basic Energy Sciences, Division of Materials Sciences and Engineering, and by UNL program development funds. Calculations were performed at the Physics Laboratory for High-Performance Computing of Renmin University of China and at the Shanghai Supercomputer Center.

**References**

1.   Tran, V.; Soklaski, R.; Liang, Y. F.; Yang, L. Layer-Controlled Band Gap and Anisotropic Excitons in Few-Layer Black Phosphorus. *Phys. Rev. B* **2014**, *89*, 235319.
2.   Xia, F.; Wang, H.; Jia, Y. Rediscovering Black Phosphorus as an Anisotropic Layered Material for Optoelectronics and Electronics. *Nat. Commun.* **2014**, *5*, 4458.
3.   Liu, H.; Neal, A. T.; Zhu, Z.; Luo, Z.; Xu, X. F.; Tomanek, D.; Ye, P. D. D. Phosphorene: An Unexplored 2D Semiconductor with a High Hole Mobility. *ACS Nano.* **2014**, *8*, 4033-4041.
4.   Li, L. K.; Yu, Y. J.; Ye, G. J.; Ge, Q. Q.; Ou, X. D.; Wu, H.; Feng, D. L.; Chen, X. H.; Zhang, Y. B. Black Phosphorus Field-Effect Transistors. *Nat. Nanotechnol.* **2014**, *9*, 372-377.
5.   Buscema, M.; Groenendijk, D. J.; Steele, G. A.; van der Zant, H. S.; Castellanos-Gomez, A. Photovoltaic Effect in Few-Layer Black Phosphorus PN Junctions Defined by Local Electrostatic Gating. *Nat. Commun.* **2014**, *5*, 4651.
6.   Qiao, J.; Kong, X.; Hu, Z. X.; Yang, F.; Ji, W. High-Mobility Transport Anisotropy and Linear Dichroism in Few-Layer Black Phosphorus. *Nat. Commun.* **2014**, *5*, 4475.
7.   Meng, L.; Wang, Y. L.; Zhang, L. Z.; Du, S. X.; Wu, R. T.; Li, L. F.; Zhang, Y.; Li, G.; Zhou, H. T.; Hofer, W. A.; Gao, H. J. Buckled Silicene Formation on Ir(111). *Nano Lett* **2013**, *13*, 685-690.
8.   Castellanos-Gomez, A.; Vicarelli, L.; Prada, E.; Island, J. O.; Narasimha-Acharya, K. L.; Blanter, S. I.; Groenendijk, D. J.; Buscema, M.; Steele, G. A.; Alvarez, J. V.; Zandbergen, H. W.; Palacios, J. J.; van der Zant, H. S. J. Isolation and Characterization of Few-Layer Black Phosphorus. *2D Mater.* **2014**, *1*, 025001.
9.   Hu, Z.-X.; Kong, X.; Jingsi, Q.; Normand, B.; Ji, W. Interlayer Electronic Hybridization Leads to Exceptional Thickness-Dependent Vibrational Properties in Few-Layer Black Phosphorus. *Nanoscale* **2016**, *8*, 2740-2750.
10.   Wang, X. M.; Jones, A. M.; Seyler, K. L.; Tran, V.; Jia, Y. C.; Zhao, H.; Wang, H.; Yang, L.; Xu, X. D.; Xia, F. N. Highly Anisotropic and Robust Excitons in Monolayer Black

Phosphorus. *Nat. Nanotechnol.* **2015**, *10*, 517-521.

11.    Keyes, R. W. The Electrical Properties of Black Phosphorus. *Phys. Rev.* **1953**, *92*, 580-584.

12.    Morita, A. Semiconducting Black Phosphorus. *Appl. Phys. A* **1986**, *39*, 227-242.

13.    Novoselov, K. S.; Jiang, D.; Schedin, F.; Booth, T. J.; Khotkevich, V. V.; Morozov, S. V.; Geim, A. K. Two-Dimensional Atomic Crystals. *Proc. Natl. Acad. Sci. USA.* **2005**, *102*, 10451-10453.

14.    Lee, C.; Yan, H.; Brus, L. E.; Heinz, T. F.; Hone, J.; Ryu, S. Anomalous Lattice Vibrations of Single- and Few-Layer $MoS_2$. *ACS Nano* **2010**, *4*, 2695-2700.

15.    Yau, S. L.; Moffat, T. P.; Bard, A. J.; Zhang, Z. W.; Lerner, M. M. STM of the (010) Surface of Orthorhombic Phosphorus. *Chem. Phys. Lett.* **1992**, *198*, 383-388.

16.    Doganov, R. A.; O'Farrell, E. C. T.; Koenig, S. P.; Yeo, Y. T.; Ziletti, A.; Carvalho, A.; Campbell, D. K.; Coker, D. F.; Watanabe, K.; Taniguchi, T.; Neto, A. H. C.; Ozyilmaz, B. Transport Properties of Pristine Few-Layer Black Phosphorus by Van Der Waals Passivation in an Inert Atmosphere. *Nat. Commun.* **2015**, *6*, 6647.

17.    Island, J. O.; Steele, G. A.; van der Zant, H. S. J.; Castellanos-Gomez, A. Environmental Instability of Few-Layer Black Phosphorus. *2D Mater.* **2015**, *2*, 011002.

18.    Ziletti, A.; Carvalho, A.; Campbell, D. K.; Coker, D. F.; Neto, A. H. C. Oxygen Defects in Phosphorene. *Phys. Rev. Lett.* **2015**, *114*, 046801.

19.    Favron, A.; Gaufres, E.; Fossard, F.; Phaneuf-L'Heureux, A. L.; Tang, N. Y. W.; Levesque, P. L.; Loiseau, A.; Leonelli, R.; Francoeur, S.; Martel, R. Photooxidation and Quantum Confinement Effects in Exfoliated Black Phosphorus. *Nat. Mater.* **2015**, *14*, 826-+.

20.    Yasaei, P.; Behranginia, A.; Foroozan, T.; Asadi, M.; Kim, K.; Khalili-Araghi, F.; Salehi-Khojin, A. Stable and Selective Humidity Sensing Using Stacked Black Phosphorus Flakes. *ACS Nano.* **2015**, *9*, 9898-9905.

21.    Kang, J.; Wells, S. A.; Wood, J. D.; Lee, J. H.; Liu, X.; Ryder, C. R.; Zhu, J.; Guest, J. R.; Husko, C. A.; Hersam, M. C. Stable Aqueous Dispersions of Optically and Electronically Active Phosphorene. *Proc. Natl. Acad. Sci. U S A* **2016**, doi/10.1073/pnas.1602215113.

22.    Huang, Y.; Sutter, E.; Shi, N. N.; Zheng, J.; Yang, T.; Englund, D.; Gao, H. J.; Sutter, P. Reliable Exfoliation of Large-Area High-Quality Flakes of Graphene and Other Two-Dimensional Materials. *ACS Nano* **2015**, *9*, 10612-20.

23.    Lim, T. B.; Polanyi, J. C.; Guo, H.; Ji, W. Surface-Mediated Chain Reaction through Dissociative Attachment. *Nat. Chem.* **2011**, *3*, 85-89.

24.    Lim, T. B.; McNab, L. R.; Polanyi, J. C.; Guo, H.; Ji, W. Multiple Pathways of Dissociative Attachment: $CH_3Br$ on Si(100)-2×1. *J. Am. Chem. Soc.* **2011**, *133*, 11534-11539.

25.    Wood, J. D.; Wells, S. A.; Jariwala, D.; Chen, K. S.; Cho, E.; Sangwan, V. K.; Liu, X. L.; Lauhon, L. J.; Marks, T. J.; Hersam, M. C. Effective Passivation of Exfoliated Black Phosphorus Transistors against Ambient Degradation. *Nano Lett.* **2014**, *14*, 6964-6970.

26.    Du, Y. L.; Ouyang, C. Y.; Shi, S. Q.; Lei, M. S. Ab Initio Studies on Atomic and Electronic Structures of Black Phosphorus. *J. Appl. Phys.* **2010**, *107*, 093718.

27.    Stull, D. R. Vapor Pressure of Pure Substances. Organic and Inorganic Compounds. *Industrial & Engineering Chem.* **1947**, *39*, 517-540.

28.    Wang, F.; Ng, W. K. H.; Yu, J. C.; Zhu, H. J.; Li, C. H.; Zhang, L.; Liu, Z. F.; Li, Q. Red Phosphorus: An Elemental Photocatalyst for Hydrogen Formation from Water. *Appl. Catal.*

*B-Environ.* **2012**, *111*, 409-414.

29.    Pecht, M.; Deng, Y. L. Electronic Device Encapsulation Using Red Phosphorus Flame Retardants. *Microelectron. Reliab.* **2006**, *46*, 53-62.

30.    Mal, P.; Breiner, B.; Rissanen, K.; Nitschke, J. R. White Phosphorus Is Air-Stable within a Self-Assembled Tetrahedral Capsule. *Science.* **2009**, *324*, 1697-1699.

31.    Chou, T. D.; Lee, T. W.; Chen, S. L.; Tung, Y. M.; Dai, N. T.; Chen, S. G.; Lee, C. H.; Chen, T. M.; Wang, H. J. The Management of White Phosphorus Burns. *Burns* **2001**, *27*, 492-497.

32.    Wang, C. G.; Huang, K.; Ji, W. Dissociative Adsorption of $CH_3X$ (X = Br and Cl) on a Silicon(100) Surface Revisited by Density Functional Theory. *J. Chem. Phys.* **2014**, *141*.

33.    Meng, S.; Zhang, Z.; Kaxiras, E. Tuning Solid Surfaces from Hydrophobic to Superhydrophilic by Submonolayer Surface Modification. *Phys. Rev. Lett.* **2006**, *97*, 036107.

34.    Bolotin, K. I.; Sikes, K. J.; Jiang, Z.; Klima, M.; Fudenberg, G.; Hone, J.; Kim, P.; Stormer, H. L. Ultrahigh Electron Mobility in Suspended Graphene. *Solid State Commun.* **2008**, *146*, 351-355.

35.    Huang, Y.; Sutter, E.; Sadowski, J. T.; Cotlet, M.; Monti, O. L. A.; Racke, D. A.; Neupane, M. R.; Wickramaratne, D.; Lake, R. K.; Parkinson, B. A.; Sutter, P. Tin Disulfide-an Emerging Layered Metal Dichalcogenide Semiconductor: Materials Properties and Device Characteristics. *ACS Nano* **2014**, *8*, 10743-10755.

36.    Li, S. L.; Wakabayashi, K.; Xu, Y.; Nakaharai, S.; Komatsu, K.; Li, W. W.; Lin, Y. F.; Aparecido-Ferreira, A.; Tsukagoshi, K. Thickness-Dependent Interfacial Coulomb Scattering in Atomically Thin Field-Effect Transistors. *Nano Lett.* **2013**, *13*, 3546-3552.

37.    Wang, L.; Sofer, Z.; Pumera, M. Voltammetry of Layered Black Phosphorus: Electrochemistry of Multilayer Phosphorene. *Chemelectrochem.* **2015**, *2*, 324-327.

38.    Prokop, M.; Bystron, T.; Bouzek, K. Electrochemistry of Phosphorous and Hypophosphorous Acid on a Pt Electrode. *Electrochimica Acta* **2015**, *160*, 214-218.

39.    Blochl, P. E. *Projector Augmented-Wave Method*. *Phys. Rev. B.* **1994**, *50*, 17953-17979.

40.    Kresse, G.; Joubert, D. From Ultrasoft Pseudopotentials to the Projector Augmented-Wave Method. *Phys. Rev. B* **1999**, *59*, 1758-1775.

41.    Kresse, G.; Furthmuller, J. Efficient Iterative Schemes for Ab Initio Total-Energy Calculations Using a Plane-Wave Basis Set. *Phys. Rev. B* **1996**, *54*, 11169-11186.

42.    Lee, K.; Murray, E. D.; Kong, L. Z.; Lundqvist, B. I.; Langreth, D. C. Higher-Accuracy Van Der Waals Density Functional. *Phys. Rev. B* **2010**, *82*, 081101.

43.    Dion, M.; Rydberg, H.; Schroder, E.; Langreth, D. C.; Lundqvist, B. I. Van Der Waals Density Functional for General Geometries. *Phys. Rev. Lett.* **2004**, *92*, 246401.

44.    Klimes, J.; Bowler, D. R.; Michaelides, A. Chemical Accuracy for the Van Der Waals Density Functional. *J. Phys.: Condens. Mat.* **2010**, *22*, 022201.

45.    Klimes, J.; Bowler, D. R.; Michaelides, A. Van Der Waals Density Functionals Applied to Solids. *Phys. Rev. B* **2011**, *83*, 195131.

46.    Heyd, J.; Scuseria, G. E.; Ernzerhof, M. Hybrid Functionals Based on a Screened Coulomb Potential. *J. Chem. Phys.* **2003**, *118*, 8207-8215.

47.    Heyd, J.; Scuseria, G. E.; Ernzerhof, M. Hybrid Functionals Based on a Screened Coulomb Potential. *J. Chem. Phys.* **2003**, *118*, 8207.

48.    Henkelman, G.; Uberuaga, B. P.; Jonsson, H. A Climbing Image Nudged Elastic Band Method for Finding Saddle Points and Minimum Energy Paths. *J. Chem. Phys.* **2000**, *113*,

9901-9904.

**Degradation of Black Phosphorus: The Role of Oxygen and Water**

Yuan Huang,[1,2,3†] Jingsi Qiao,[4†] Kai He,[2] Stoyan Bliznakov,[5] Eli Sutter,[6] Xianjue Chen,[1,3] Da Luo,[1,3] Fanke Meng,[5] Dong Su,[2] Jeremy Decker,[1] Wei Ji,[4,*] Rodney S. Ruoff,[1,3*] and Peter Sutter[7,*]

[1] Center for Multidimensional Carbon Materials (CMCM), Institute for Basic Science (IBS), Ulsan 689-798, Republic of Korea

[2] Center for Functional Nanomaterials, Brookhaven National Laboratory, Upton, NY 11973, USA

[3] Department of Chemistry and School of Materials Science and Engineering, Ulsan National Institute of Science and Technology, Ulsan 689-798, Republic of Korea

[4] Department of Physics and Beijing Key Laboratory of Optoelectronic Functional Materials & Micro-nano Devices, Renmin University of China, Beijing 100872, China

[5] Chemistry Department, Brookhaven National Laboratory, Upton, New York 11973, USA

[6] Department of Mechanical & Materials Engineering, University of Nebraska-Lincoln, Lincoln, NE 68588, USA

[7] Department of Electrical & Computer Engineering, University of Nebraska-Lincoln, Lincoln, NE 68588, USA

†These authors contributed equally to this work. *To whom correspondence should be addressed: Peter Sutter (E-mail: psutter@unl.edu); Rodney S. Ruoff (E-mail: ruofflab@gmail.com); Wei Ji (E-mail: wji@ruc.edu.cn).

## Contents:

## S1. Sample preparation and microscopic characterization

Black phosphorus flakes were exfoliated onto $SiO_2$/Si substrates, followed by optical microscopy imaging. The substrates were cleaned using acetone, isopropanol and de-ionized (DI) water, and then $O_2$ plasma was used for further cleaning. After putting adhesive tape with black phosphorus flakes in contact with the $SiO_2$/Si substrate, the substrate was annealed in air at 80 °C for 1 min on a conventional laboratory hot plate. We then cooled down the substrate to room temperature, and finally peeled off the tape to complete the exfoliation process. This exfoliation method could improve the yield ratio of the flakes with different thicknesses, as described in detail in our previous work.[1] After exfoliation, the flakes were characterized using different methods and conditions as described in the main text.

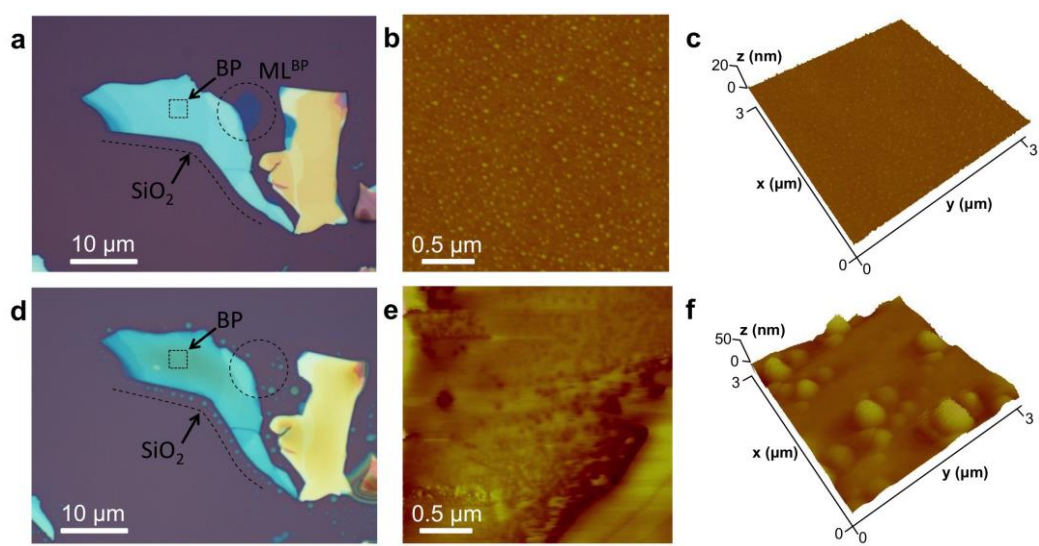

**Figure S1. Optical and AFM images of an exfoliated black phosphorus flake supported by $SiO_2$/Si. a.** Optical image of the freshly exfoliated, pristine black phosphorus flake. **b.** AFM image within the flake, obtained in the area outlined by a dashed square in a. **c.** AFM image of the surface of the monolayer black phosphorus flake (ML[BP], marked by a dashed circle in a). **d.** Optical image after exposing the sample to air for 1 day. Note the buildup of a chain of islands/droplets along the flake edge (e.g., dashed line), and the disappearance of an entire monolayer black phosphorus segment (marked ML[BP] in a, and outlined with a dashed circle). **e.** AFM image within the flake (dashed square). **f.** AFM image at the former location of the monolayer black phosphorus flake, showing increased corrugation in the form of large islands/droplets.

Figure S1 shows optical and AFM images of a pristine black phosphorus flake, and of the same flake after exposure to air for 1 day. Within 1 hour after exfoliation, the surface of the flake appeared optically flat and the surrounding $SiO_2$ was featureless in optical microscopy (Figure S1a). At this stage, AFM showed arrays of small clusters decorating the black phosphorus surface (Figure S1b). The same droplet morphology was found on the surface of thick, bulk-like (Figure S1b) and monolayer flakes (Figure S1c). After ambient exposure for 1 day, several changes are observed. Optical microscopy detects relatively minor changes to thick black phosphorus flakes. However, their outline now appears more rounded, and the surrounding $SiO_2$ support shows large, drop-like features in the vicinity of all flakes (Figure S1d). The surface of bulk-like flakes is no longer flat but now appears rough. The well-defined small droplets seen shortly after exfoliation are no longer clearly observable (Figure S1e), but seem to have coalesced to a thin film after the small clusters grew larger. The thinnest portions (e.g., the monolayer segment, $ML^{BP}$ marked by a circle in Figure S1a) are completely decomposed and only large drops remain in their place on the substrate (Figure S1f).

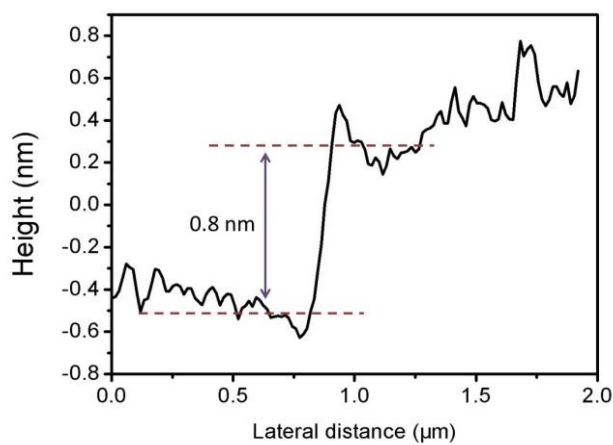

**Figure S2.** AFM height profile of monolayer BP. The measured thickness is about 0.8 nm.

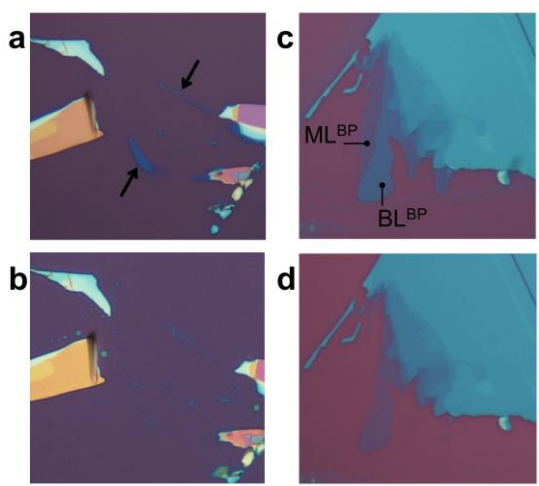

**Figure S3. Effects of air exposure and water immersion on exfoliated black phosphorus flakes. a.** Optical micrograph of freshly exfoliated black phosphorus flakes supported on $SiO_2$/Si. **b.** Same sample region after exposure to air for 2 days. **c.** Freshly exfoliated black phosphorus flakes. **d.** Same sample region after immersion in DI water for 2 days.

Figure S3a, c are optical images of pristine BP flakes, captured within 5 min after exfoliation onto a SiO$_2$/Si substrate. The second set of images (Figure S3a, c) shows the same sample areas after 2 days of exposure to air. Thin flakes (Figure S3, arrows) were completely etched away. 2 days of immersion in DI water (with dissolved O$_2$ at equilibrium with air) completely removed monolayer flakes, whereas bilayer areas were transformed into monolayer (Figure S3d). The comparison with the sample exposed to air shows that decomposition of black phosphorus is substantially slower in DI water, but some residual reactivity remained.

## S2. Setup for gas bubbling

To probe the reactivity of black phosphorus in contact with DI water with controlled $O_2$ content (achieved by bubbling either $N_2$ or $O_2$ to deplete or enrich dissolved oxygen, respectively), exfoliated black phosphorus flakes on $SiO_2/Si$ substrates were put in a glass container filled with DI water (Figure S4). A long needle was immersed into the DI water for gas ($N_2$ or $O_2$) bubbling, and the outlet tube (left) was fed in another beaker with DI water. The flow rate could be controlled by valves, but this parameter is not critical for our experiments.

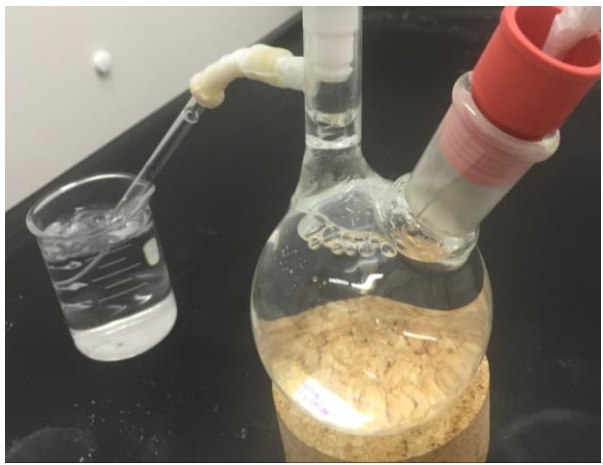

**Figure S4. Photograph of the setup used for probing the stability of black phosphorus in DI water with gas ($N_2$ or $O_2$) bubbling.** Black phosphorus flakes on $SiO_2/Si$ substrate was held in the glass container, in which DI water could either be depleted or enriched in dissolved $O_2$.

## S3. Electrical transport measurement

### 1. Electrical performance of BP FET devices in air

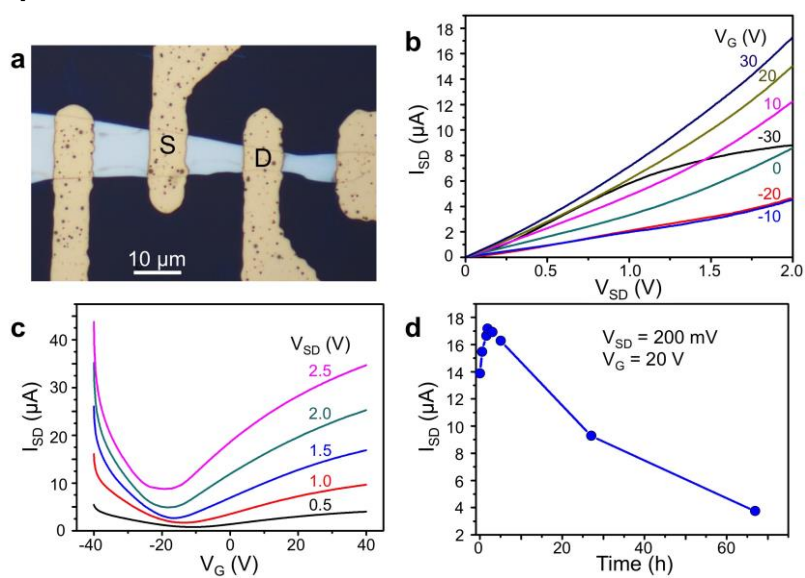

**Figure S5. Electrical transport in a back-gated black phosphorus FET device (on SiO₂/Si) exposed to air. a.** Optical image of the black phosphorus FET device. S: source, D: drain electrode. **b.** $I_{SD}$-$V_{SD}$ curves for $V_{SD}$ ranging from 0 to 2 V and back-gate voltage $V_G$ ranging from -30 V to 30 V. **c.** Ambipolar relationship between $I_{SD}$ and $V_G$ for $V_{SD}$ ranging from 0.5 to 2.5 V. **d.** $I_{SD}$ as a function of time of exposure to ambient air.

Figure S5a shows the overall configuration of our devices (channel thickness: ~15 nm), which were fabricated on SiO₂/Si substrates with Ti (5 nm)/Au (50 nm) as source and drain contact electrodes. Figure S5b,c summarize measurements on a representative device, performed using three-terminal DC field-effect transistor characteristics with back gating via the SiO₂ dielectric at room temperature in ambient air. At low source-drain voltage ($V_{SD}$ from 0 - 0.5 V), the current-voltage ($I_{SD}$ -$V_{SD}$) characteristics are linear over the entire range of back-gate voltages from -30 to +30 V, which indicates ideal ohmic contacts between the Ti/Au electrodes and the BP channel. As the back-gate voltage $V_G$ is changed from -30 V to +30 V, the drain current, $I_{SD}$, first decreases and then increases again, confirming the ambipolar transfer characteristics shown in previous reports.[2] This ambipolar behavior is also clearly seen in Figure S5c, which shows $I_{SD}$-$V_G$ characteristics for $V_G$ from -40 V to +40 V, with source-drain voltage varying from 0.5 to 2.5 V in steps of 0.5 V. The field effect mobility extracted from these characteristics is ~230 cm²/V·s for holes and ~30 cm²/V·s for electrons, respectively, consistent with mobility values observed previously.[3] We then tested the stability of another BP FET device in air. The results are summarized by plotting values of $I_{SD}$ at constant conditions ($V_{SD}$ = 200 mV; $V_G$ = +20 V) as a function of air exposure time (Figure S5d). The conductance of the device initially increases, as shown by an increase of $I_{SD}$ from 14 μA to 17 μA in the first 5 hours. A saturation of $I_{SD}$ is followed by a continuous decrease to ~4 μA over the next 60 hours. The initial rise in conductance may be explained by an effect of current annealing,[4] which could lower the contact resistance between the black phosphorus channel and the source- and drain metal electrodes. After long time expose to air, our results (discussed in the main text) suggest that the black phosphorus flake becomes oxidized, which we see here as a decrease in the channel conductance. This effect of degradation of our black phosphorus FET device is similar to that reported previously for unencapsulated devices, which also showed an initial increase in $I_{SD}$ followed by a decrease of the channel current, ultimately leading to the complete breakdown of the device after ~56 hours in air.[5]

## 2. Capacitance of the DI water top gate of a black phosphorus field-effect transistor (FET) device

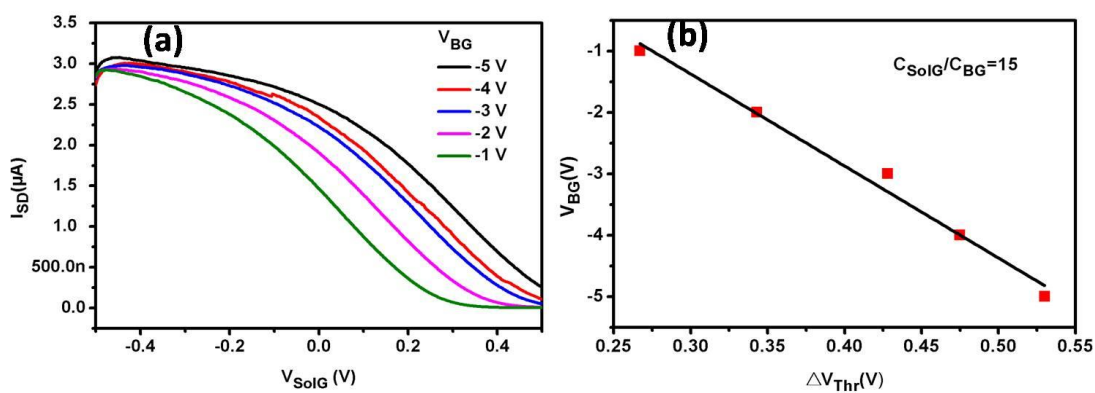

**Figure S6. Measurement of the capacitance of the DI water top gate of a BP FET device. (a)** Transfer characteristics ($I_{SD}$ *vs.* $V_{SolG}$) for different back gate voltages ($V_{BG}$ from -1 V to -5 V). **(b)** Plot of the back gate voltage *vs.* shift in threshold voltage ($\Delta V_{Thr}$) of the transfer curves, showing a linear dependence from which the ratio between the back gate and solution top gate capacitances can be derived.

The capacitance of the DI water solution gate is determined using a lever-arm principle by measuring the change in the transfer characteristics of the solution gated device ($I_{SD}$ *vs.* $V_{SolG}$) due to different applied back gate bias voltages, $V_{BG}$ (Figure S6 (a)). Back gating induces well-defined shifts in the transfer characteristic, which can be quantified by a linear fit to determine the shift in the threshold voltage, $\Delta V_{Thr}$, for different values of $V_{BG}$. From the linear dependence shown in Figure S6b, the solution gate capacitance $C_{SolG}$ can be determined *via* $C_{SolG}/C_{BG} = \Delta V_{BG}/\Delta V_{Thr}$.[6, 7] Using the known specific capacitance of the 300 nm $SiO_2$ dielectric of the back gate, $C_{BG} = 11.6$ nF/cm$^2$,[8] we determine a specific capacitance of the DI water solution top gate in contact with a black phosphorus FET channel of 174 nF/cm$^2$. Using this value, the data shown in Figure. S6(a) translate into a room temperature mobility of 1260 cm$^2$/V.s for the solution gated black phosphorus FET.

## S4. Stability of black phosphorus flakes in acidic and basic solutions

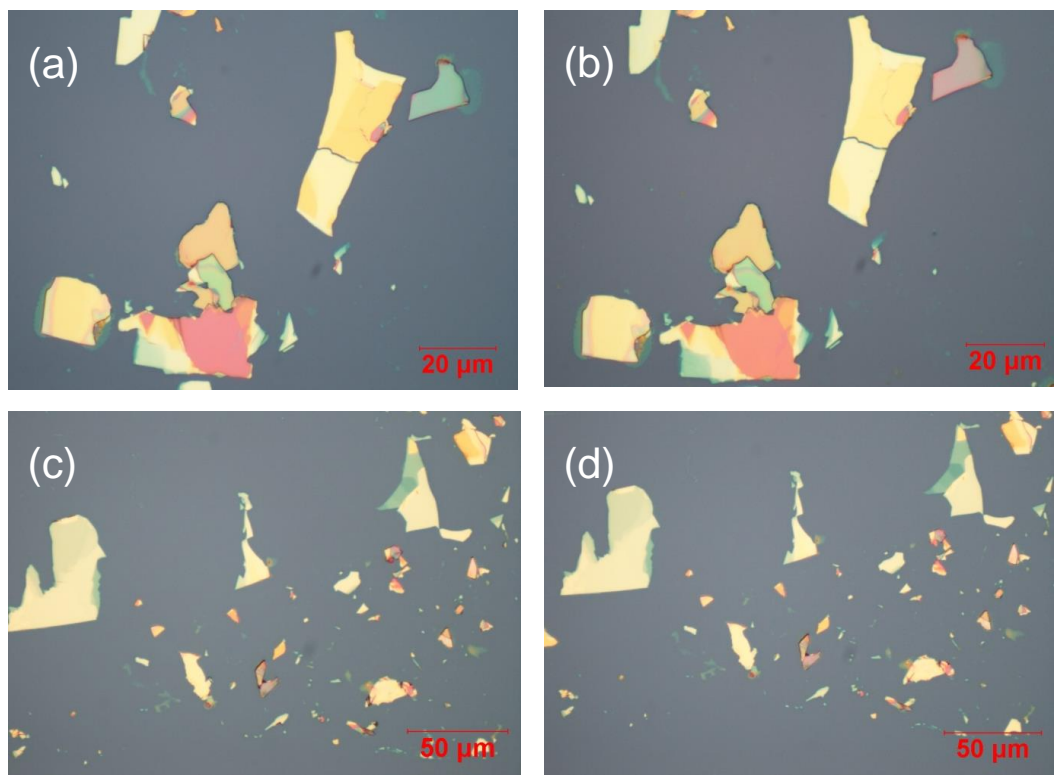

**Figure S7. Black phosphorus in acidic and basic solutions. a.** and **c.** are optical images of freshly exfoliated black phosphorus flakes on SiO$_2$/Si substrates. **b.** The same black phosphorus flakes shown in (a) after immersion in 37% aqueous hydrochloric acid (HCl) for 30 minutes. **d.** The same black phosphorus shown in (c) after immersion in 2M potassium hydroxide (KOH) for 30 minutes.

To test the stability of black phosphorus in strong acids and bases, we immersed the exfoliated BP flakes in 37 % HCl and 2M KOH. After 30 minutes immersion in these strong acids and bases, the flakes did not show any obvious change, as can be seen in Fig. S7b,d. After immersion for 2 h, we inspected the flakes again by optical microscopy. The flake exposed to HCl solution still did not show any significant change, i.e., remained nearly unchanged as shown in Fig. S7 a, b. But some of the flakes in KOH solution disappeared after 1 h. We assign this to the etching of SiO$_2$ by KOH, but not a reaction of KOH with black phosphorus.

## S5. Electrochemistry on black phosphorus

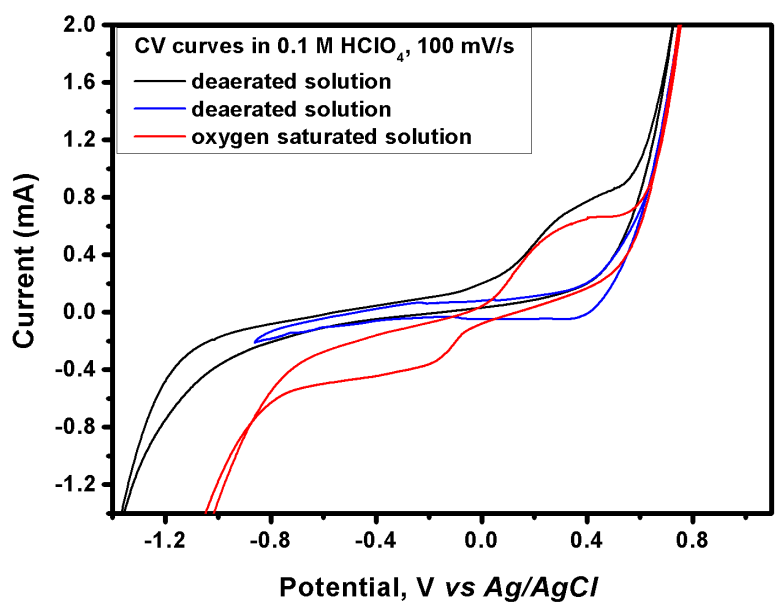

**Figure S8 – Cyclic voltammograms (CVs) measured on a black phosphorus working electrode in 0.1M perchloric acid (HClO4) solution (Ag/AgCl reference electrode).** Comparison of CVs obtained in oxygen saturated (red curve) and $N_2$ deaerated solutions (black, blue curves). In the blue curve, the potential, V, was restricted to small negative potentials below the threshold for the hydrogen evolution reaction (HER). Scan rate: 100 mV/s. Area of the black phosphorus sample: 50 mm[2].

Representative cyclic voltammograms (CVs) of a BP electrode measured in oxygen-saturated and deaerated 0.1 M $HClO_4$ solutions are compared in Figure S8. A broad oxidation peak at ~ +0.4 V is observed at similar position and intensity in both cases, *i.e.*, it obviously does not depend on the concentration of dissolved oxygen in the electrolyte. This peak corresponds to the electrochemical oxidation of elemental phosphorus ($P^0$) to a higher oxidation state.

This suggests that a $P_xO_y$ film is formed on the surface of the black phosphorus in this potential range. At more positive potentials (higher than +0.5 V), the phosphorus is further oxidizing to $P^{5+}$, forming ($PO_4^{3-}$) ions that dissolve into the solution and form the end product, phosphoric acid ($H_3PO_4$). These observations are consistent with our EELS results, as well as recently reported results by Wang *et al.*[9] The oxidation peak at +0.4 V is not observed in deaerated solution (shown in blue in Figure S8) if the reverse potential sweep is stopped before reaching the threshold for the hydrogen evolution reaction (HER, -0.9 V). This means that hydrogen gas generated at potentials more negative than -1.0 V chemically reduces the $P_xO_y$ on the electrode surface to elemental phosphorus.[10] Thus, when the potential is swept in the anodic direction, the bare phosphorus surface (not covered by an oxide) is oxidized, and the peak at +0.4 V is observed. Finally, we also observe a cathodic peak at -0.2 V, which appears only in the CV curve measured in oxygen-saturated solution. We attribute this peak to the oxygen reduction reaction (ORR) taking place on the black phosphorus electrode surface in this potential range.

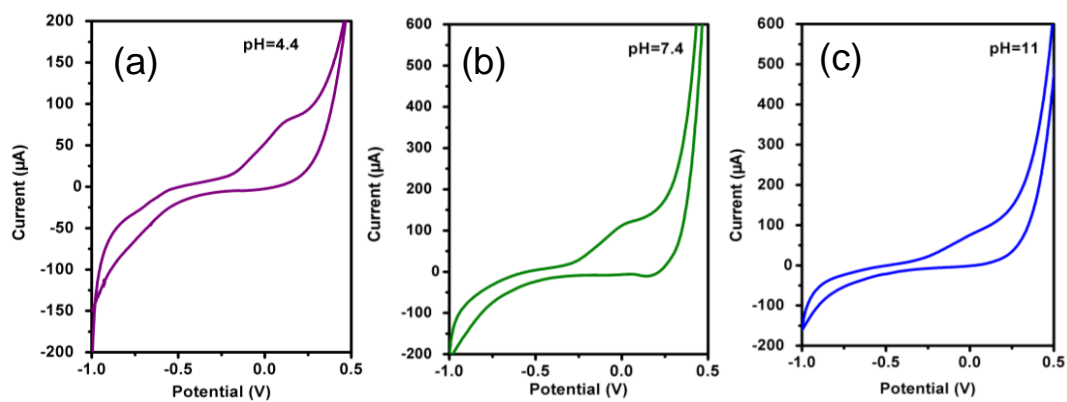

**Figure S9. Cyclic voltammograms (CVs) measured on a black phosphorus working electrodes in solutions with different pH (relative to Ag/AgCl reference electrode).** Comparison of CVs obtained in different pH value solutions (4.4, 7.4 and 11) with potential range from -1 V to 0.5 V. The solutions are: **(a)** $NaH_2PO_4$ (pH = 4.4); **(b)** $CH_3COONa+CH_3COOH$ (pH = 7.4); **(c)** $Na_3PO_4+Na_2HPO_4$ (pH = 11). No gas bubbling during these measurements. Scan rate: 100 mV/s. Area of the black phosphorus sample: 50 mm$^2$.

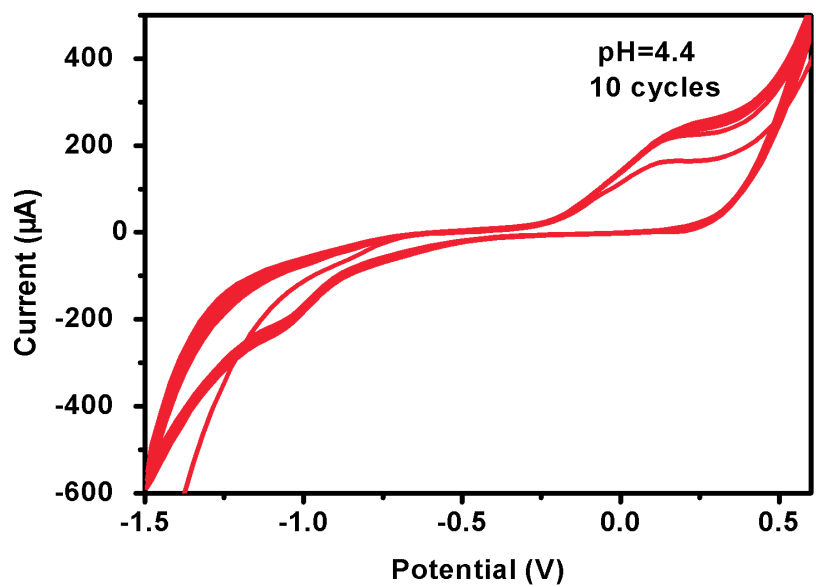

**Figure S10. Cyclic voltammograms (CVs) measured on a black phosphorus working electrode in $NaH_2PO_4$ (pH = 4.4) solution (Ag/AgCl reference electrode.** The measurement were repeated for 10 cycles in the potential range of -1.5 V to 0.5 V. No gas bubbling was observed during this measurement. Scan rate: 100 mV/s. Area of the black phosphorus sample: 50 mm$^2$.

## S6. DFT calculations

### 1. Geometric structures and bandstructures of O/BP.

A single O atom absorbed on BP has five adsorption sites: Lone pair O (LP-O), interstitial bridge O (IB-O), dimer bridge O (DiB-O), horizontal bridge O (HB-O) and diagonal bridge O (DgB-O). The structural stability is LP-O/BP > IB-O/BP > DiB-O/BP > DgB-O/BP > HB-O/BP. Single O atoms not only induce the geometric structures of BP in Fig. S11, but also bring changes of the effective mass and impurities states for the bandstructures of pristine BP shown in Fig. S12.

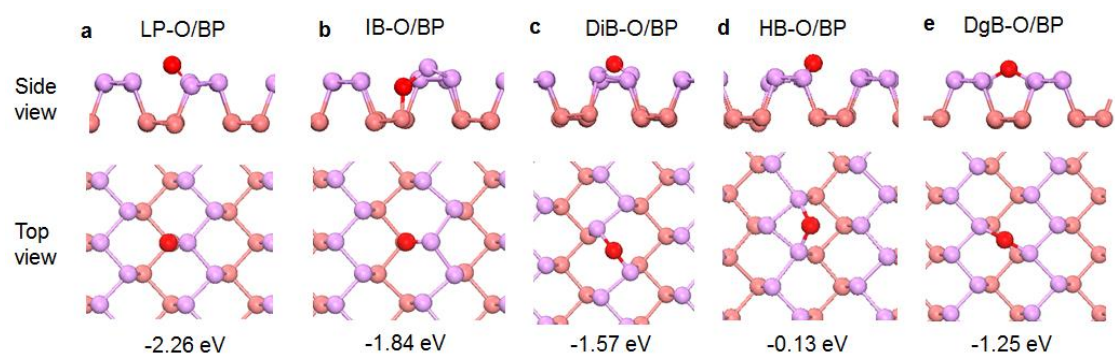

**Figure S11.** Configurations of an O atom adsorbed on 1L BP. **a.** Lone pair (LP-). **b.** Interstitial bridge (IB-). **c.** Dimer bridge (DiB-). **d.** Horizontal bridge (HB-). **e.** Diagonal bridge (DgB-).

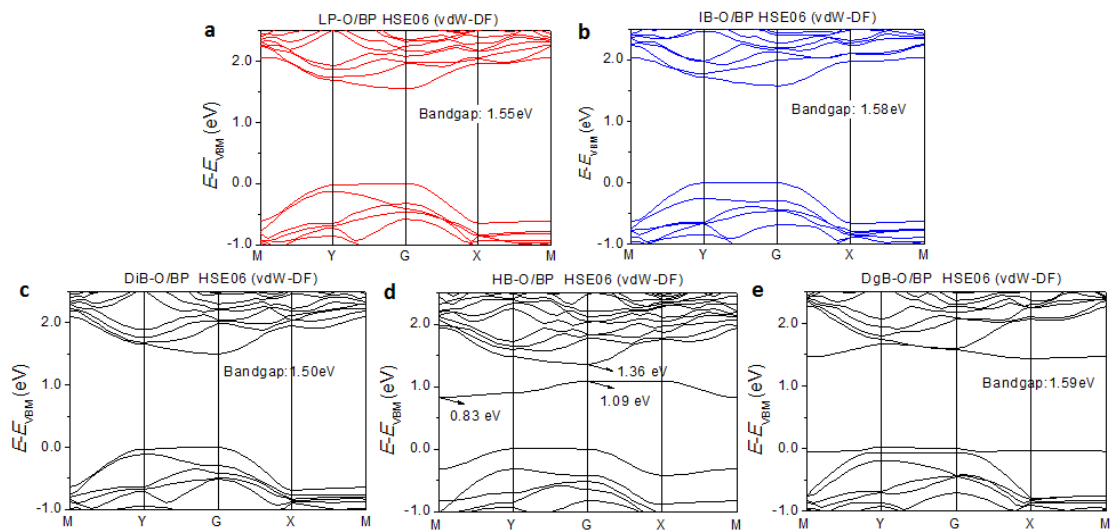

**Figure S12.** Band structures of lone pair (L-) (a), interstitial bridge (I-) (b), dimer bridge (DI-) (c), horizon bridge (HB-) (d) and diagonal bridge (DG-) (e) O atom adsorbed on 1L BP.

## 2. 25 types of initial adsorption structures of H₂O/BP.

Different adsorption sites of O atom and H atom, different orientations of $H_2O$ and different initial adsorption distances are considered to confirm the most stable structure of $H_2O/BP$ (in Fig. 3e). The detailed information is shown in Fig. S13.

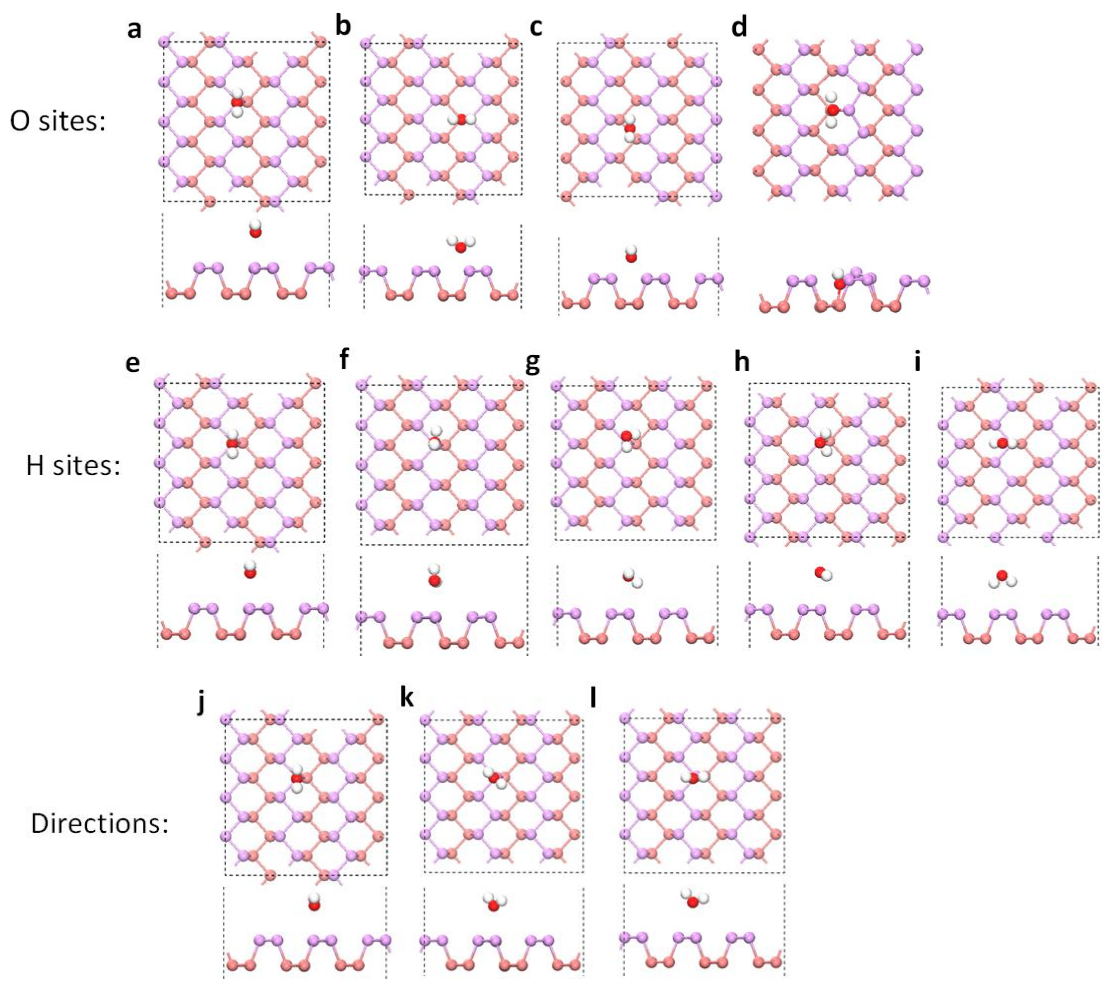

**Figure S13. Initial structures of H₂O/BP. Different adsorption sites of O atom:** above the upper P (**a**), lone pair O (**b**), diagonal bridge O (**c**) and interstitial bridge O (**d**). **Different orientations of H atoms:** two H atoms up (**e**), one up one parallel (**f**), one down one parallel (**g**), two parallel (**h**), two down (**i**). Different directions of $H_2O$ (**j-l**).

## S7. Black phosphorus dispersed in DI water

Deionized water was freeze pump thawed in a round bottom flask five times using a Schlenk line and a dry ice/acetone bath. The water was frozen, then exposed to vacuum. The flask was sealed, and the water allowed to degas. This procedure was followed five times to maximize the removal of any dissolved gasses (namely oxygen). Via a glove bag, under nitrogen atmosphere, 5 mg of black phosorus crystal was introduced to 500 ml of the freeze pumped water. This was then ultrasonicated for 16 hours in an ice bath under nitrogen atmosphere. As the sonication proceeded, the black phosphorus crystals slowly dispersed into a gray and opalescent dispersion. With further sonication, it became a turbid gray suspension as shown in Fig. S14a. This suspension was centrifuged for 10 minutes at 5000 rpm and the supernatant was used for subsequent analysis, and no obvious color change within 4 months.

For transmission electron microscopy, lacey carbon coated copper grids were utilized. For optical and Raman imaging 300 nm $SiO_2$/Si substrates were used. A drop of the centrifuged solutions was transferred, in a glove bag under nitrogen atmosphere, onto the $SiO_2$/Si and the TEM grids in a Schlenk vial. The water was removed under vacuum and the samples stored under nitrogen. Any transfers from the sealed glassware to instrumentation were carried out as rapidly as possible to reduce degradation from light and atmosphere.

A JEM-2100 $LaB_6$ TEM was used to image bright field images at 200 kV. Raman spectra were measured using a micro Raman Spectrometer (Alpha 300s, WITec GmbH) with a 532 nm laser. Data were collected for 20 s at ~0.1 mW using a 100X objective. Raman measurements were carried out under argon to prevent oxidative degradation.

Figure S14b shows an optical micrograph of a BP flake which was analyzed by the Raman spectrometer (Fig. S14d) to show the characteristic spectra of BP with Raman bands at 361 $cm^{-1}$, 438 $cm^{-1}$ and 465 $cm^{-1}$, corresponding to the $Ag^1$, $B_{2g}$ and $Ag^2$ modes of black phosphorus. No thickness measurements were performed as yet, so no claims towards particle thickness can be reliably made. However, the TEM micrograph in Fig. S14c does seem to indicate that the black phosphorus particles removed from the water suspension may indeed be quite thin, and indeed show little evidence of pitting or other degradation, despite long term processing in water under aggressive conditions.

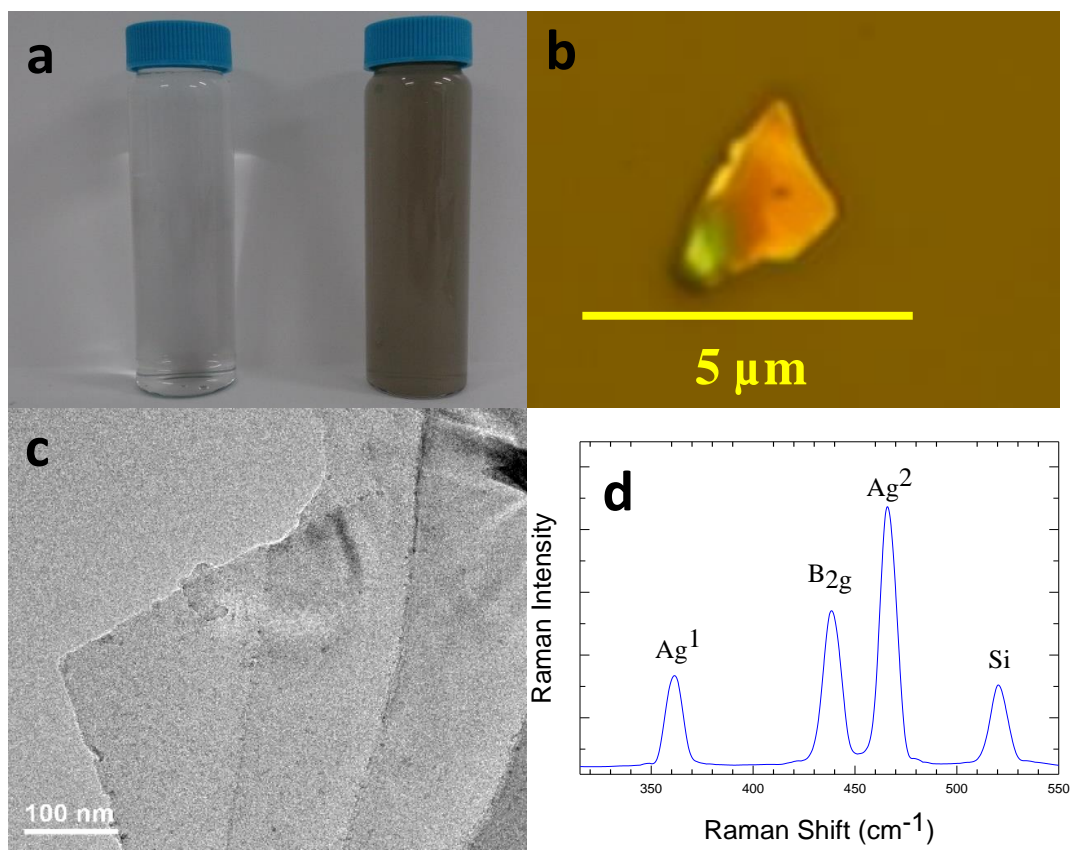

**Figure S14. a.** Black phosphorus dispersion after 16 hours ultrasonication. **b.** Optical image of freshly solution-exfoliated black phosphorus flakes on SiO$_2$/Si substrates. **c.** Transmission electron micrograph of BP on lacey carbon film. **D**. Raman spectra of BP flake.

## S8. Oxygen isotope labeling and TOF-SIMS measurement

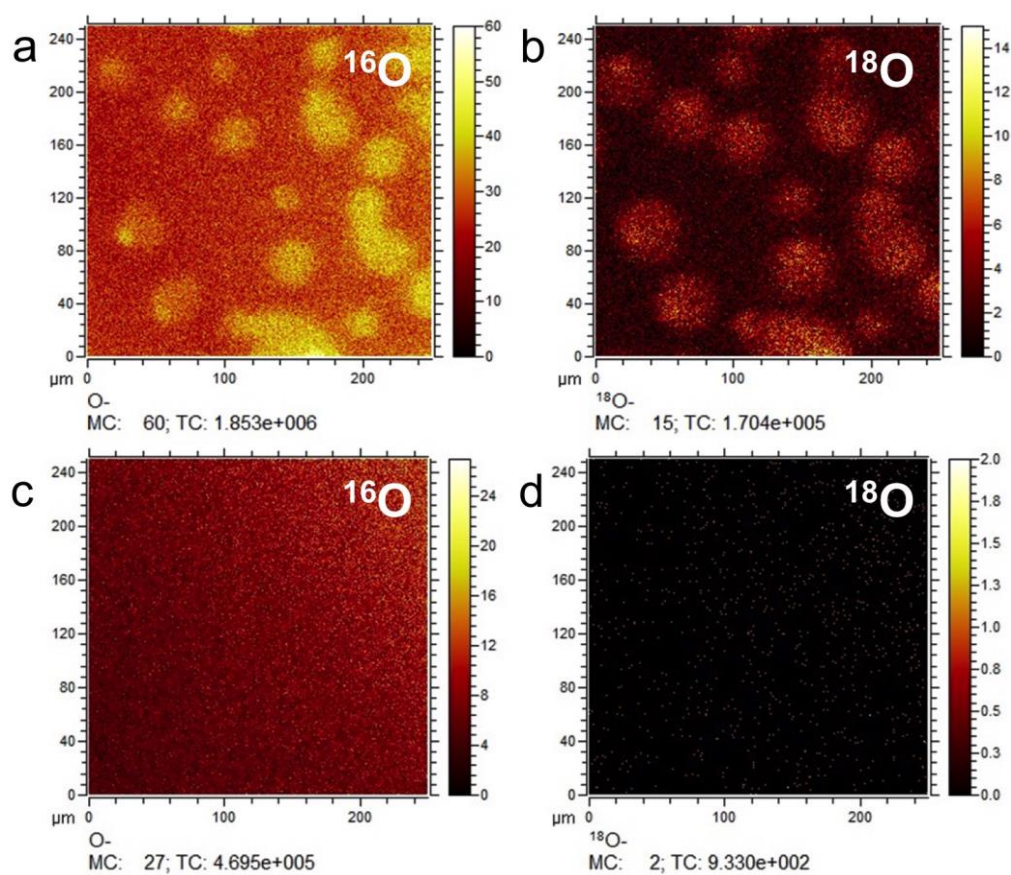

**Figure S15.** Time-of-Flight Secondary Ion Mass Spectrometry (TOF-SIMS) mapping for BP flakes. **a.** $^{16}O$ map, and **b.** $^{18}O$ map of a BP flake exposed to $^{18}O_2/H_2{}^{16}O$ in a closed cell for one day. **c.** $^{16}O$ map, and **d.** $^{18}O$ map for a freshly cleaved BP flake. MC: Maximum ion count in one pixel; TC: Total ion count in the entire image. Color scale legends: Ion count.

To confirm the dominant role of dissolved oxygen in the oxidative attack of black phosphorus in aqueous solutions, we carried out an isotope labeling experiment in which BP flakes were exposed to a mixture of isotopically pure $^{18}O_2$ and water ($H_2{}^{16}O$) vapor. Black phosphorus flakes peeled off from a bulk crystal were put in a closed flask containing $^{18}O_2$ and saturated vapor above a reservoir of liquid $H_2{}^{16}O$. Prior to introducing the BP, dissolved $^{16}O_2$ in $H_2{}^{16}O$ was first removed by de-aeration using the $N_2$ bubbling method described in the manuscript and in SI Section S2. Subsequently, the isotope labeled oxygen gas was added by several cycles of evacuation of the flask, followed by loading with $^{18}O_2$ to atmospheric pressure. This procedure ensured a mixed $^{18}O_2/H_2{}^{16}O$ environment with minimal residual $^{16}O_2$ background.

Time-of-Flight Secondary Ion Mass Spectrometry (TOF-SIMS) was performed after one day of exposure of the BP flakes to the $^{18}O_2/H_2{}^{16}O$ mixture, and on a freshly cleaved BP flake for comparison. Some oxidation (by $^{16}O_2$) during the initial handling and transfer in air is responsible for a uniform $^{16}$O background present in both the sample exposed to $^{18}$O$_2$/H$_2^{16}$O and the freshly cleaved reference sample (with similar intensity). In addition, the sample exposed to the $^{18}$O$_2$/H$_2^{16}$O mixture shows patches with higher $^{18}$O signal, which coincide with regions of increased $^{16}$O count (Fig. S15 a, b). This result suggests that the sample was additionally oxidized by the $^{18}$O$_2$ in the flask, and that the H$_2^{16}$O in turn selectively adsorbed in these more strongly oxidized areas. This behavior is consistent with the DFT calculations showing that the native (i.e., freshly cleaved) BP surface is hydrophobic but turns increasingly hydrophilic with progressive oxidization. From the TOF-SIMS maps, we can infer that the absorbed water was limited to the oxidized regions (where $^{18}$O is present) because the surroundings of these regions are hydrophobic. The appearance of the water adsorption regions is similar to the droplets observed by optical microscopy and AFM.

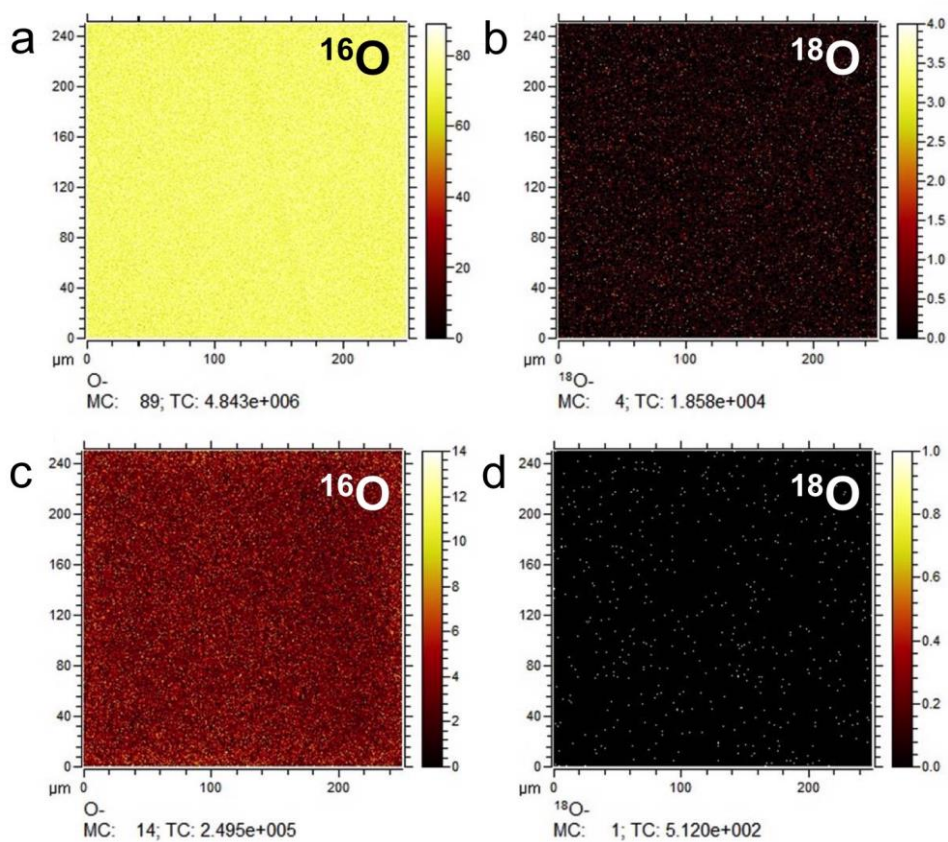

**Figure S16.** TOF-SIMS mapping images of two BP flakes. **a.** $^{16}$O map, and **b.** $^{18}$O map of a BP flake that was exposed to air for one day. **c.** $^{16}$O map, and d. $^{18}$O map of a BP flake exposed to H$_2^{18}$O saturated vapor for one day. MC: Maximum ion count in one pixel; TC: Total ion count in the entire image. Color scale legends: Ion count.

In a second isotope labeling experiment, we compared the effects of air and of saturated water vapor on the black phosphorus surface. BP flakes exposed to air for one day showed a very high $^{16}$O signal, as can be seen in Fig. S16a. The abundance of

$^{18}$O in air is only 0.204%, so the signal from $^{18}$O is much smaller but it is still detectable in the air-exposed BP flakes (Fig. S16b). For comparison, we exposed a second BP flake for one day to saturated vapor above a reservoir of isotopically labeled, deaerated (i.e., $^{16}O_2$ depleted) water ($H_2^{18}O$). As can be seen in Fig. S16c,d, the $^{18}$O signal measured on this sample remained very low, which implies that the surface was not oxidized and that no detectable water absorbed on BP surface. This result further supports our conclusions that BP is not oxidized by exposure to water and that the pristine BP surface is hydrophobic.

## S9. Supplementary References

1.  Huang, Y.; Sutter, E.; Shi, N. N.; Zheng, J. B.; Yang, T. Z.; Englund, D.; Gao, H. J.; Sutter, P. Reliable Exfoliation of Large-Area High-Quality Flakes of Graphene and Other Two-Dimensional Materials. *ACS Nano* **2015**, *9*, 10612-10620.

2.  Li, L. K.; Yu, Y. J.; Ye, G. J.; Ge, Q. Q.; Ou, X. D.; Wu, H.; Feng, D. L.; Chen, X. H.; Zhang, Y. B. Black Phosphorus Field-Effect Transistors. *Nat. Nanotechnol.* **2014**, *9*, 372-377.

3.  Liu, H.; Neal, A. T.; Zhu, Z.; Luo, Z.; Xu, X. F.; Tomanek, D.; Ye, P. D. D. Phosphorene: An Unexplored 2d Semiconductor with a High Hole Mobility. *ACS Nano.* **2014**, *8*, 4033-4041.

4.  Bolotin, K. I.; Sikes, K. J.; Jiang, Z.; Klima, M.; Fudenberg, G.; Hone, J.; Kim, P.; Stormer, H. L. Ultrahigh Electron Mobility in Suspended Graphene. *Solid State Commun.* **2008**, *146*, 351-355.

5.  Wood, J. D.; Wells, S. A.; Jariwala, D.; Chen, K. S.; Cho, E.; Sangwan, V. K.; Liu, X. L.; Lauhon, L. J.; Marks, T. J.; Hersam, M. C. Effective Passivation of Exfoliated Black Phosphorus Transistors against Ambient Degradation. *Nano Lett.* **2014**, *14*, 6964-6970.

6.  Perera, M. M.; Lin, M. W.; Chuang, H. J.; Chamlagain, B. P.; Wang, C. Y.; Tan, X. B.; Cheng, M. M. C.; Tomanek, D.; Zhou, Z. X. Improved Carrier Mobility in Few-Layer MoS$_2$ Field-Effect Transistors with Ionic-Liquid Gating. *ACS Nano* **2013**, *7*, 4449-4458.

7.  Huang, Y.; Sutter, E.; Sadowski, J. T.; Cotlet, M.; Monti, O. L. A.; Racke, D. A.; Neupane, M. R.; Wickramaratne, D.; Lake, R. K.; Parkinson, B. A.; Sutter, P. Tin Disulfide—an Emerging Layered Metal Dichalcogenide Semiconductor: Materials Properties and Device Characteristics. *ACS Nano* **2014**, *8*, 10743-10755.

8.  Farmer, D. B.; Lin, Y. M.; Avouris, P. Graphene Field-Effect Transistors with Self-Aligned Gates. *Appl. Phys. Lett.* **2010**, *97*, 013103.

9.  Wang, L.; Sofer, Z.; Pumera, M. Voltammetry of Layered Black Phosphorus: Electrochemistry of Multilayer Phosphorene. *Chemelectrochem* **2015**, *2*, 324-327.

10. Prokop, M.; Bystron, T.; Bouzek, K. Electrochemistry of Phosphorous and Hypophosphorous Acid on a Pt Electrode. *Electrochimica Acta* **2015**, *160*, 214-218.